\documentclass[aps,pra,twocolumn,showpacs,superscriptaddress,longbibliography]{revtex4-1} 
\usepackage{graphicx}
\usepackage{amsmath}
\usepackage{amssymb}
\usepackage{epstopdf}
\usepackage{amsfonts}
\usepackage{dcolumn}
\usepackage{multirow}

\begin{document}
\title{Tunneling dynamics of two interacting one-dimensional particles}
\author{Seyed Ebrahim Gharashi}
\affiliation{Department of Physics and Astronomy,
Washington State University,
  Pullman, Washington 99164-2814, USA}
\author{D. Blume}
\affiliation{Department of Physics and Astronomy,
Washington State University,
  Pullman, Washington 99164-2814, USA}
\date{\today}

\begin{abstract}
We present 
one-dimensional
simulation results 
for
the cold atom tunneling experiments
by the Heidelberg group
[G. Z\"urn {\em{et al.}}, Phys. Rev. Lett. {\bf{108}}, 075303 (2012) and 
G. Z\"urn {\em{et al.}}, Phys. Rev. Lett. {\bf{111}}, 175302 (2013)] 
on one or two $^6$Li atoms confined by a potential 
that consists of an approximately harmonic optical trap plus a linear 
magnetic field gradient.
At the non-interacting 
particle level, we find that the WKB (Wentzel-Kramers-Brillouin)
approximation may not be used as a reliable tool to extract the 
trapping potential parameters from the experimentally measured tunneling data.
We use our numerical calculations along with the experimental tunneling rates 
for the non-interacting system to reparameterize the trapping potential.
The reparameterized trapping potentials serve as input for our 
simulations of two interacting particles.
For two interacting (distinguishable) atoms on the upper branch, we
reproduce the experimentally measured tunneling rates,
which vary over several orders of magnitude, 
fairly well. 
For infinitely strong interaction strength, 
we compare the time dynamics with that of two identical fermions and 
discuss the implications of fermionization on the dynamics.
For two attractively-interacting atoms on the molecular branch, we find
that 
single-particle tunneling dominates for weakly-attractive interactions
while pair tunneling dominates for strongly-attractive interactions.
Our first set of calculations yields
qualitative but not quantitative agreement 
with the experimentally measured tunneling rates. 
We obtain quantitative agreement 
with the experimentally measured tunneling rates if we
allow for a weakened radial confinement. 
\end{abstract}
\maketitle

\section{Introduction}
\label{sec_introduction}

Open quantum systems are at the heart of many physical phenomena from 
nuclear physics to quantum information theory~\cite{Breuer-OQS, Wiseman-OQS}.
In fact, all ``real'' quantum systems are, to some extent, open systems.
Interactions with the environment cause decoherence, resulting in
non-equilibrium dynamics.
It is often simpler to design experiments that probe non-equilibrium physics 
than it is to design experiments that probe equilibrium physics.
Conversely, the theoretical toolkit for describing systems in equilibrium 
is generally much farther developed than that for describing systems in 
non-equilibrium.

Ultracold atom systems provide a platform for realizing clean and tunable 
quantum systems~\cite{BlochReview, BlumeReview, GiorginiReview, ChinReview}. 
Over the past few years, much effort has gone into describing 
non-equilibrium experiments that are accessible, within approximate 
or exact frameworks, to theory. Notable experiments are the equilibration 
dynamics of one-dimensional Bose gases~\cite{Langen15}, the spin 
dynamics of dipolar molecules in optical lattices with low filling 
factor~\cite{Yan13}, and the tunneling dynamics of effectively 
one-dimensional few-fermion systems~\cite{Jochim2,Jochim4}.
This paper focuses on the latter set of experiments. 
Specifically, the goal of the present work is to describe the 
tunneling dynamics of few-fermion systems, which are prepared in a 
well defined quasi-eigenstate (metastable state), into free space. 
We consider small systems and directly 
solve the time-dependent Schr\"odinger equation in coordinate space.
As we will show, this approach provides a means to quantify the 
importance of the particle-particle interaction, covering time 
scales from a fraction of the trap scale to thousands times the trap scale.
Alternatively, one could adopt a quantum optics perspective
and pursue a master equation approach.

Tunneling is arguably the most quantum phenomenon there is: If the system was behaving classically, tunneling would be absent~\cite{Razavi13}. Tunneling plays an important role across physics, chemistry and technology. The scanning tunneling microscope~\cite{Oka14}, for example, nicely illustrates how a physics phenomenon, the tunneling of electrons, has been turned into a powerful practical tool (the imaging of materials). The $\alpha$-decay, i.e., the decay of a $^4$He nucleus from a heavy nucleus, is an example discussed in most undergraduate physics texts (see,
for example, Ref.~\cite{Serway04}). The typical picture is to identify an effective reaction coordinate and to obtain the tunneling rate from a 
WKB analysis. While powerful, such treatments completely neglect the effect of interactions. Interactions also play a crucial role in sorting out under which conditions electrons in light atoms tunnel sequentially or simultaneously~\cite{Eckle08}. 
The 
two-particle system considered in this work has been realized experimentally and is the possibly simplest scenario that deals with a truly open quantum system (the atoms can escape to infinity) in which interactions (short-range atom-atom interactions) play a crucial role. As we will show, even for this relatively simple set-up, matching theory and experiment is a non-trivial task. Of course, two-particle tunneling has been investigated previously in 
this and related 
contexts~\cite{Lode12, Hunn13, Kim11, Lode09,Rontani12,Rontani13,Lundmark15}.

The remainder of this paper is organized as follows.
Section~\ref{sec_system} introduces the Hamiltonian,
the Heidelberg experiment and selected simulation details.
Sections~\ref{sec_molecular} and \ref{sec_upper}
discuss the molecular and upper branch tunneling dynamics.
For both cases, it is argued that the trapping potential needs
to be reparameterized.
Using the reparameterized trapping potential, 
numerical simulations for the 
tunneling dynamics of two distinguishable $^6$Li atoms on the molecular 
branch  
and the upper branch
are 
discussed.
Comparisons with the experimentally measured tunneling rates are presented.
Finally, Sec.~\ref{sec_summary} summarizes and provides an outlook.
Simulation details and some technical 
aspects 
are relegated to 
Appendices~\ref{app_BR}--\ref{app_flux-anal}.


\section{System Hamiltonian and simulation details}
\label{sec_system}

\subsection{One-body Hamiltonian, WKB analysis, and Heidelberg experiment}
\label{subsec_single-Ham}

\begin{figure}[htbp]
\centering
\includegraphics[angle=0,width=0.4\textwidth, clip=true]{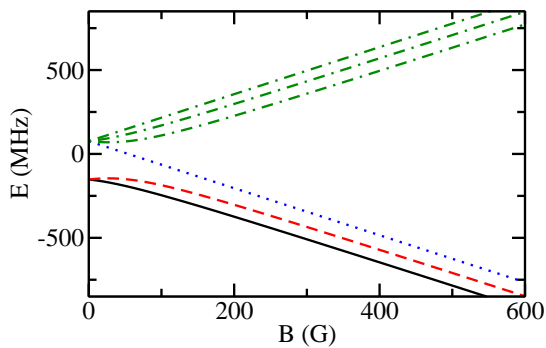}
\caption{(Color online)
Energy of the hyperfine states of $^6$Li as a function of the magnetic field
strength $B$. Solid, dashed and dotted lines correspond to states
$| 1 \rangle$,
$| 2 \rangle$, and
$| 3 \rangle$,
respectively (see text for details).  
States
$| 1 \rangle$ and
$| 2 \rangle$ are used in the upper branch experiments~\cite{Jochim2}, 
while states
$| 1 \rangle$ and
$| 3 \rangle$
are used in the molecular branch experiments~\cite{Jochim4}.
The higher-lying energy states shown by dash-dotted lines are not relevant 
for the present paper.
}
\label{fig_BR-energy}
\end{figure}

\begin{figure}[htbp]
\centering
\includegraphics[angle=0,width=0.4\textwidth, clip=true]{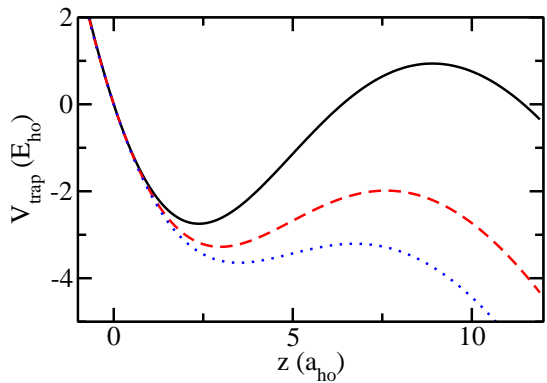}
\caption{(Color online) The trapping potential, Eq.~(\ref{Vtrap}),
for $\mathcal{C} = 1892$G/m
and three different values of $p$, 
$p=0.795$ (solid line), $p=0.6875$ (dashed line), and $p=0.63496$
(dotted line). 
$V_0$ and $z_{\text{R}}$ are fixed at the values reported in Table~\ref{table-exp}.
}
\label{fig_trap}
\end{figure}

This section considers a single $^6$Li atom 
with mass $m$.
The atom is assumed to be in the hyperfine state $| F, m_F \rangle$. 
We consider the three lowest  hyperfine states of the $^6$Li atom, 
referred to as
$| 1 \rangle = | 1/2,1/2 \rangle$,
$| 2 \rangle = | 1/2,-1/2 \rangle$, and
$| 3 \rangle = | 3/2,-3/2 \rangle$.
Figure~\ref{fig_BR-energy}
shows the dependence of the hyperfine 
energy levels on the magnetic field strength $B$.
The atom with coordinates $(x,y,z)$ is trapped optically in
a non-separable potential that is much tighter in the $\rho$-direction
($\rho^2 = x^2+y^2$) than in the $z$-direction~\cite{Jochim2,Jochim4}.
Throughout this work, we do not simulate the motion in the tight transverse
confining direction.
The transverse trapping frequency does, however, enter
into the calculation of the renormalized one-dimensional coupling constant 
(see Sec.~\ref{subsec_two-body_Ham}).
Evaluating the confinement created by the gaussian laser
beam at $\rho=0$, the effective one-dimensional
single-particle Hamiltonian $H^{\text{sp}}$ reads~\cite{Jochim2,Jochim4}
\begin{eqnarray}
H^{\text{sp}} (z,t; p, z_{\text{R}}, \mathcal{C}_{|j \rangle}(B)) = 
-\frac{\hbar^2}{2m} \frac{\partial ^2}{\partial z^2} \nonumber \\
+ V_{\text{trap}}(z,t; p, z_{\text{R}}, \mathcal{C}_{|j \rangle}(B)),
\label{Ham_sp}
\end{eqnarray}
where
the trapping potential $V_{\text{trap}}$ along the $z$-direction 
depends implicitly on the internal or hyperfine state $|j \rangle$ of the 
atom through the coefficient $\mathcal{C}_{|j \rangle}$,
\begin{eqnarray}
V_{\text{trap}}(z,t; p, z_{\text{R}}, \mathcal{C}_{|j \rangle}(B)) = \nonumber \\
p(t) V_0 \left(1 - \frac{1}{\left(z/z_{\text{R}} \right)^2 + 1} \right) - 
\mu_\text{B} \mathcal{C}_{|j\rangle}(B) z.
\label{Vtrap}
\end{eqnarray} 
The first term on the right hand side of Eq.~(\ref{Vtrap}) accounts for the 
optical confinement.
$V_0$ denotes the maximum depth of the trap, $p(t)$ 
a time-dependent parameter
[$p(t) \le 1$], and
$z_{\text{R}}$ the Rayleigh range of the laser beam that produces the confinement.
The second term on the right hand side of Eq.~(\ref{Vtrap}) is linear in $z$
and makes the tunneling possible.
$\mu_\text{B}$ is the Bohr magneton and 
$\mathcal{C}_{|j\rangle}(B)$ depends on the hyperfine state, 
magnetic field strength
and magnetic field gradient $B'$,
\begin{eqnarray}
\mathcal{C}_{|j\rangle}(B) = c_{|j\rangle}(B) B'.
\label{cfac}
\end{eqnarray} 
Here, $c_{|j\rangle}(B)$ is a dimensionless parameter close to $1$ 
(see below for details).
Table~\ref{table-exp}
\begin{table}
\caption{Parameters from Refs.~\cite{Jochim2, Jochim4} that define the 
trapping potential.
Since the energy of the two-particle system on the
molecular branch is smaller than
the energy of the two-particle system on the
upper branch,
the $p(t=0)$ value for the molecular branch
is chosen to be smaller than that for the upper branch; 
this guarantees that the tunneling rates for the two experiments have
roughly comparable orders of magnitude.
The harmonic oscillator units are defined in terms of 
$\omega = 2 \pi \times 1234$Hz, corresponding to 
$E_{\text{ho}} = 8.177 \times 10^{-31}$J,
$a_{\text{ho}} = 1.167 \mu $m,
and
$\omega^{-1} = 1.290 \times 10^{-4}$s,
or
$1 \text{J} = 1.223 \times 10^{30} E_{\text{ho}}$,
$1 \text{m} = 8.570 \times 10^{5}  a_{\text{ho}}$, and
$1 \text{s} = 7753 \omega^{-1}$.
In an alternative levitation measurement, the magnetic field gradient 
was found to be $B' = 1890(20)$G/m~\cite{Jochim2}.
}
\begin{tabular}{c|c}
Quantity & 
Value
\\

\hline
\hline
$V_0$ & 
$k_B \times 3.326 \mu \text{K} = 56.16 E_{\text{ho}}$~~
\\

$z_{\text{R}}$ & 
$9.975(5) \times 10^{-6} \text{m} = 8.548(5) a_{\text{ho}}$~~
\\  

$p(-t_r)$&
$0.795$
\\  

$p(t=0)$ (upper branch)&
$0.6875$
\\  

$p(t = 0)$ (molecular branch)&
$0.63496$
\\  

$dp/dt$ (for $-t_r<t<0$)&
$-43 \text{s}^{-1}$
\\

$ B'$ (WKB approximation)~~& 
$1892$G/m
\end{tabular}
\label{table-exp}
\end{table}
summarizes the trap parameters reported by the Heidelberg 
group~\cite{Jochim2, Jochim4};
the parameters are obtained from a combination of measurement and WKB 
analysis. 
Figure~\ref{fig_trap}
shows $V_{\text{trap}}$ for
$\mathcal{C} = 1892$G/m
and three different values of $p$.
The solid line shows the typical confinement at the
beginning of the experiment while the dashed and dotted lines show typical 
confinements during the hold time of the upper branch
and molecular branch experiments, respectively (see below for details).

Since the trapping potential changes with time, there exists no
set of units that characterizes the system
equally well for all times.
Throughout, following Ref.~\cite{Jochim2}, we choose 
$\omega = 2 \pi \times 1234$Hz to define the oscillator units
$E_{\text{ho}}$, $a_{\text{ho}}$, and $T_{\text{ho}}$: 
$E_{\text{ho}} = \hbar \omega$, $a_{\text{ho}} = \sqrt{\hbar / (m \omega)}$,
and $T_{\text{ho}} = 2 \pi / \omega$.

The confining potential $V_{\text{trap}}$ has a local minimum at 
$z_{\text{min}}$ 
and a local maximum at $z_b$.
To gain insight into the harmonic approximation, we expand
$V_{\text{trap}}$ around its local minimum and calculate the frequency 
$\omega_{\text{trap}}(p)$ of the harmonic term,
\begin{eqnarray}
\omega_{\text{trap}}(p) = \sqrt{2 \frac {p(t) V_0}{m} 
\frac
{\left(z_{\text{R}}^4 - 3 z_{\text{R}}^2  z_{\text{min}}^2 \right)}
{\left( z_{\text{R}}^2 + z_{\text{min}}^2\right)^3}
}.
\label{om-trap}
\end{eqnarray} 
In the absence of the magnetic field gradient $B'$, the minimum 
of $V_{\text{trap}}$ is located at 
$z_{\text{min}}=0$. 
For a finite magnetic field gradient, the local minimum 
$z_{\text{min}}$ depends on the parameters of the trapping potential.
The frequency $\omega_{\text{trap}}(p)$ can differ notably from 
the frequency $\omega$
and provides, in some cases, a more natural unit.
We define
$E_{\text{trap}}(p) = \hbar \omega_{\text{trap}}(p)$,
$a_{\text{trap}}(p) = \sqrt{\hbar /(m \omega_{\text{trap}}(p))}$, and
$T_{\text{trap}}(p) = 2 \pi / \omega_{\text{trap}}(p)$.
Note that these units depend explicitly on $p(t)$; correspondingly, we
specify $p(t)$ when we use these units.

\begin{figure}[htbp]
\centering
\includegraphics[angle=0,width=0.4\textwidth, clip=true]{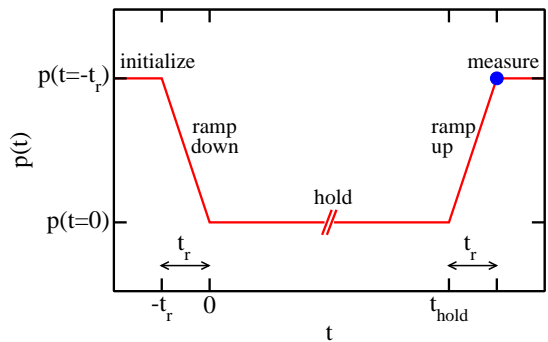}
\caption{(Color online) 
Schematic of the time sequence of the experiment.
After initialization of the system, the dimensionless
parameter $p(t)$ decreases from
$p(t=-t_r)$ to $p(t=0)$ with a rate $dp/dt=-43$ s$^{-1}$,
remains constant for $t_{\text{hold}}$ ($t_{\text{hold}} \gg t_r$), and increases
to its initial value over the time period $t_r$.
The measurement is performed at the time $t_{\text{hold}}+t_r$.
}
\label{fig_p}
\end{figure}

The single-particle tunneling dynamics is, to a good approximation,
described by an exponential decay,
\begin{eqnarray}
P_{\text{sp,in}}(t) = 
P_{\text{sp,in}}(t_{\text{ref}})
\exp [- \gamma_{\text{sp}} (t - t_{\text{ref}})],
\label{eq_exp-decay}
\end{eqnarray} 
where $P_{\text{sp,in}}(t)$ denotes the probability of finding the particle 
inside the trap,
the tunneling rate $\gamma_{\text{sp}}$ is assumed to be 
constant, and $t_{\text{ref}}$ is a reference time.
Within the WKB approximation (see, e.g., 
Ref.~\cite{Ankerhold07}), 
the tunneling rate $\gamma_{\text{sp}}^{\text{WKB}}$ reads
\begin{eqnarray}
\label{eq_WKB-rate}
\gamma_{\text{sp}}^{\text{WKB}} = f^{\text{WKB}} {\mathcal{T}},
\end{eqnarray}
where the frequency $f^{\text{WKB}}$ 
and the tunneling coefficient
${\mathcal{T}}$ are 
 given by
\begin{eqnarray}
\label{eq_WKB-rate1}
f^{\text{WKB}} = 
\frac {\epsilon-V_{\text{trap}}(z_{\text{min},t=0})} {2 \pi \hbar}
\end{eqnarray}
and 
\begin{eqnarray}
\label{eq_WKB-rate2}
{\mathcal{T}} =
\exp \left(-2 \int_{z_{\epsilon,2}}^{z_{\epsilon,3}}
\sqrt{\frac {2 m} {\hbar^2} |\epsilon - V_{\text{trap}}(z)|} d z \right).
\end{eqnarray}
In Eqs.~(\ref{eq_WKB-rate1})-(\ref{eq_WKB-rate2}),
$V_{\text{trap}}$ is the trapping potential with $p(t=0)$ 
(see Fig.~\ref{fig_p} for the time dependence of $p$),
$z_{\text{min},t=0}$ is the $z$-value at which $V_{\text{trap}}$ 
with $p(t=0)$
takes its local minimum, and 
the WKB energy $\epsilon$ of state $n$ is found by the consistency condition
\begin{eqnarray}
\int_{z_{\epsilon,1}}^{z_{\epsilon,2}} 
\sqrt{2 m [\epsilon - V_{\text{trap}}(z)]} d z = 
\left(n + \frac {1} {2} \right) \pi \hbar.
\label{eq_WKB-en}
\end{eqnarray} 
Here,  $z_{\epsilon,1}$,  $z_{\epsilon,2}$, and $z_{\epsilon,3}$
with $z_{\epsilon,1} < z_{\epsilon,2} < z_{\epsilon,3}$ are the 
three solutions 
of $\epsilon - V_{\text{trap}}(z)=0$ and $n$ with $n=0,1,2, \dots$ 
denotes the order of the semiclassical ``bound state'' of the trap.
In theory, one has $P_{\text{sp,in}}(t) + P_{\text{sp,out}}(t) = 1$,
with the initial condition $P_{\text{sp,in}}(-t_r) = 1$.
Here 
$P_{\text{sp,out}}(t)$ 
denotes the probability that the particle
has left the trap.
The inside and outside regions are defined through 
$z < z_b$ and $z > z_b$, respectively, with $z_b$ corresponding to 
the barrier position at time $t = -t_r$.

We now briefly review the experimental sequence employed by the Heidelberg 
group~\cite{Jochim2,Jochim4}. 
The experiment prepared the atom in an ``eigenstate'' of the deep trap 
($p=0.795$ at $t = -t_r$) and then lowered the barrier by decreasing $p(t)$
over a time period $t_r$. At time $t=0$, $p(t)$ reached its minimum.
After a variable hold time $t_{\text{hold}}$, the barrier was ramped back up 
over a time period $t_r$. 
At time $t=t_{\text{hold}}+t_r$, the experiment monitored 
the fraction $P_{\text{sp,out}}(t)$ of the particle that had left the trap.
To obtain $P_{\text{sp,out}}(t)$, the experiment was repeated many times 
for each $t=t_{\text{hold}}+t_r$ and the
data were averaged [each individual experiment yields 
$P_{\text{sp,out}}(t_{\text{hold}}+t_r) = 0$ or 1].
The time sequence is sketched in Fig.~\ref{fig_p}.
In the experiment~\cite{Jochim4},
the initial condition was $P_{\text{sp,in}}(-t_r) < 1$
due to non-unit state preparation fidelity. 
While this changes the overall normalization, 
it does not change the tunneling dynamics.

The coefficients $c_{|j \rangle}(B)$, and correspondingly the  
$\mathcal{C}_{|j\rangle}(B)$, depend
on the magnetic field strength $B$, which is used to tune the atom-atom 
scattering length.
The coefficients $c_{| j \rangle}(B)$ can, at least in a first analysis, be
obtained using the Breit-Rabi formula~\cite{Breit-Rabi} 
(see Appendix~\ref{app_BR}).
For state $| 3 \rangle$, the Breit-Rabi coefficient 
$c_{| 3 \rangle}^{\text{BR}}(B)$ is independent of the magnetic field. 
For states $| 1 \rangle$ and $| 2 \rangle$, 
the dependence of the Breit-Rabi coefficients on the magnetic field strength 
$B$ is comparatively strong when $B$ is small ($B \lesssim 600$G) 
and weak when $B \to \infty$ ($B \gtrsim 600$G).
References~\cite{Jochim2, Jochim4} did not use the Breit-Rabi formula to 
determine the $c_{| j \rangle}(B)$ coefficients (see below for details).

To parameterize $V_{\text{trap}}$, Refs.~\cite{Jochim2, Jochim4} fed the result
from ``calibration measurements'' into
Eqs.~(\ref{eq_WKB-rate}) and~(\ref{eq_WKB-en}).
In a first step, the parameters $V_0$ and $z_{\text{R}}$ of the optical trap, 
which is independent of the hyperfine state and magnetic field
strength, 
were calibrated 
assuming $p=1$.
Specifically, the single-particle trap energy levels 
of the pure optical trap 
[${\mathcal{C}}_{| j \rangle}(B)=0$ in Eq.~(\ref{Vtrap})]
were measured 
spectroscopically and the parameters $V_0$ and $z_{\text{R}}$ were chosen 
such that the WKB energy levels agreed with the measured 
energies
(see the supplemental material of Ref.~\cite{Jochim2}).

For the upper branch tunneling experiment, $p(t = -t_r)$ and $p(t = 0)$ 
were obtained by measuring the relative integrated light intensities of the
trap beams, i.e., $p(t = -t_r)$ and $p(t = 0)$ were calibrated relative to 
$p=1$~\cite{Zurnemail2}.
To obtain $B'$, 
tunneling experiments at various magnetic fields 
using $^6$Li in state $| 2 \rangle$
were performed~\cite{footnote7a}.
To prepare the atom in an excited trap state, the experiments used a trick.
Two
atoms 
in the same hyperfine states 
were prepared in the trap
(these atoms do not interact),
forcing the two-particle system to sit in a superposition of the lowest
and first excited trap states. 
The assumption was then that the tunneling dynamics proceeds as if there
were a single particle in the first excited trap state and another single 
particle in the lowest trap state.
The tunneling was attributed to the particle in the first excited trap state
while the particle in the lowest trap state was assumed to have no chance of 
tunneling.
This assumption is, as our simulations show,
justified 
quite well
(see Appendix~\ref{app_sup-vs-pro}).
To analyze the tunneling data, $c_{|j\rangle}$ was assumed to be equal 
to 1 
for all magnetic field strengths
and 
$B'$ was adjusted to yield a WKB tunneling rate
$\gamma_{\text{sp}}^{\text{WKB}}$ that
agreed with the measured tunneling rate $\gamma_{\text{sp}}^{\text{exp}}$.
The resulting $B'$ was then used for all 
hyperfine 
states.

The two-particle molecular branch experiments were conducted at 
magnetic field strengths
varying from $350$G to $1202$G and utilized states 
$| 1 \rangle$ and $| 3 \rangle$~\cite{Jochim4}.
The parameters $p(t = -t_r)$, $V_0$, $z_{\text{R}}$, and $B'$ were taken as
those obtained from the upper branch experiments.
Compared to the upper branch experiments, $p(t=0)$ was reduced to obtain
tunneling times smaller than a few thousand 
milliseconds
and
the magnetic field dependence of the coefficients $c_{| j \rangle}(B)$
was found to play a non-negligible role.
For technical reasons, $p(t=0)$ was not calibrated via a ``direct''
photodetector measurement~\cite{Zurnemail2}.
Instead, $p(t=0)$ and $c_{| j \rangle}(B)$ were determined based on the WKB 
analysis of the experimentally measured single-particle tunneling 
rates (see supplemental material of Ref.~\cite{Jochim4}).
Specifically, the single-particle tunneling measurements were performed at 
$B=350$G and $569$G and the 
parameters $p(t=0)$ and $c_{| j \rangle}(B)$ were adjusted to yield a WKB value
$\gamma_{\text{sp}}^{\text{WKB}}$ that agreed with the measured tunneling rate
$\gamma_{\text{sp}}^{\text{exp}}$ at both $B$-fields
(see supplemental material of Ref.~\cite{Jochim4}).
The analysis yielded $p(t=0)= 0.63496$~\cite{Jochim4}. 
The $c_{| 1 \rangle}(B)$ and $c_{| 3 \rangle}(B)$ values are given in 
Table~\ref{table-sp-1}.

\subsection{Simulation of single-particle tunneling dynamics}

To determine the single-particle tunneling rate theoretically, we prepare the initial state 
($t \le -t_r$) through imaginary time propagation. The initial state can be 
thought of as a quasi-eigenstate. We then propagate the initial state in real 
time for $t > -t_r$. For $-t_r < t< 0$, we change $p(t)$ according to 
$dp/dt = -43 \text{s}^{-1}$. For $t > 0$, $p(t)$ is kept constant, i.e., 
$p(t) = p(0)$. By analyzing the flux through $z = z_b$, we calculate 
$P_{\text{sp,in}}(t)$ and $P_{\text{sp,out}}(t)$. We do not simulate the 
up ramp, i.e., the time period $t_{\text{hold}} < t < t_{\text{hold}} + t_r$,
since we found that the populations 
$P_{\text{sp,in}}(t)$ and $P_{\text{sp,out}}(t)$ do not change appreciably 
during the up-ramp. 
The simulation details are described in Appendices~\ref{app_Cheb}
and~\ref{app_im-time}.

\subsection{Two-body Hamiltonian and simulation of two-particle tunneling dynamics}
\label{subsec_two-body_Ham}

This section considers two $^6$Li atoms, each described by the 
single-particle 
Hamiltonian $H^{\text{sp}}$ [see Eq.~(\ref{Ham_sp})], that interact 
through the short-range potential $V_{\text{int}}(z_{12})$, where $z_{12} = z_1-z_2$.
The two-body Hamiltonian $H$ reads
\begin{eqnarray}
H(z_1, z_2, t; p, z_{\text{R}}, \mathcal{C}_{|j_1 \rangle}(B), \mathcal{C}_{|j_2 \rangle}(B))
= \nonumber \\
H^{\text{sp}} (z_1, t; p, z_{\text{R}}, \mathcal{C}_{|j_1 \rangle}(B)) + \nonumber\\
H^{\text{sp}} (z_2, t; p, z_{\text{R}}, \mathcal{C}_{|j_2 \rangle}(B)) 
+ V_{\text{int}}(z_{12}).
\label{eq_H-2b}
\end{eqnarray}
Since the range of the true $^6$Li-$^6$Li van der Waals potential is,
for the experiments considered, much 
smaller than the de Broglie wavelength of the atoms, the details of the 
interaction potential are not probed and the true interaction potential can be 
replaced by a simpler model potential that has the same three-dimensional 
$s$-wave scattering length $a_{\text{3D}}$ as the true atom-atom potential.
For $^6$Li the most precise 
magnetic field dependence of $a_{\text{3D}}$ is given in 
Ref.~\cite{Jochim5}.
To convert $a_{\text{3D}}$ to the one-dimensional coupling constant 
$g_{1\text{D}}$, we assume a three-dimensional zero-range potential and 
strictly harmonic confinement with angular frequency 
$\omega_{\rho}$
in the tight direction.
Describing the two-body interaction potential along the
$z$-direction by
\begin{eqnarray}
V_{\text{ZR}} (z_{12}) = g_{1\text{D}} \delta(z_{12}),
\label{eq_VZR}
\end{eqnarray}
the renormalized one-dimensional 
coupling constant $g_{1{\text{D}}}$ 
is 
given by~\cite{Olsh98}
\begin{eqnarray}
\frac{g_{1\text{D}}}{\hbar \omega_{\rho} a_{\rho}} = 
\frac{2a_{\text{3D}}}{a_{\rho}} 
\left( 
1- \frac{|\zeta(1/2)|}{\sqrt{2}} \frac{a_{\text{3D}}}{a_{\rho}}
\right)^{-1},
\label{eq_olsh}
\end{eqnarray}
where $\zeta(1/2)$ is equal to $-1.46035$ and
$a_{\rho}$ denotes the harmonic oscillator length in the tight confining 
direction, $a_{\rho} = \sqrt{\hbar / (m \omega_{\rho})}$.
The one-dimensional coupling constant $g_{\text{1D}}$ and the
one-dimensional scattering length
$a_{1\text{D}}$ are related via
$a_{1\text{D}}=-2 \hbar^2 / (m g_{1\text{D}})$.
To determine $\omega_{\rho}$,
Ref.~\cite{sala13}
analyzed the optical 
single-particle trap with $p(t)=1$ in the absence of the magnetic
field gradient, accounting for the longitudinal (weak) and transverse 
(tight)
directions.
The harmonic frequency $\omega_{\rho}$ in the transverse direction 
was found to be 
$\omega_{\rho}^{\text{ref}} = 2 \pi \times 14.22(35)$kHz~\cite{sala13}.
For $p(t) \ne 1$, $\omega_{\rho}^{\text{ref}}$ needs
to be multiplied by
$\sqrt{p(t)}$, i.e.,
$\omega_{\rho} = \sqrt{p(t)} \omega_{\rho}^{\text{ref}}$~\cite{Jochim2,Jochim4,sala13},
resulting in a time-dependent
$g_{\text{1D}}$.
As discussed at the beginning of Sec.~\ref{sec_mol-two}, the time dependence
of $g_{\text{1D}}$ has a negligible
affect on the tunneling rate and we neglect it
for the calculations presented in
Secs.~\ref{sec_mol-two} and \ref{sec_upper-two}.

The addition of the linear term [second term on the 
right hand side of Eq.~(\ref{Vtrap})] moves
the atoms away from the origin to positive $z$ values.
Using Eq.~(3) of the supplemental material of Ref.~\cite{sala13}
to model the confinement created by the
gaussian beam in the longitudinal and transverse directions
and expanding around 
$\rho=0$, one finds
that the harmonic frequency in the transverse direction
decreases with increasing $z$.
For $z=z_{\text{min}}$
($z = z_b$), we find that the harmonic frequency
in the $\rho$-direction decreases by around 14\% (38\%)
and 11\% (44\%) for the molecular
and upper branches, respectively, compared
to the frequencies for $z=0$.
This suggests that the tight confinement length
$a_{\rho}$ in the presence of the magnetic field gradient
may be larger
than $[p(t=0)]^{-1/4} a_{\rho}^{\text{ref}}$, where 
$a_{\text{ref}}=\sqrt{\hbar/(m \omega_{\rho}^{\text{ref}})}$,
and correspondingly that the coupling constant $g_{1\text{D}}$
is modified.
We return to this aspect in Secs.~\ref{sec_mol-two} and 
\ref{sec_upper-two}.
We reemphasize that the renormalization
prescription given in Eq.~(\ref{eq_olsh}) relies
on the harmonicity of the confinement.
It is well documented in the literature that 
this renormalization
prescription is modified by 
anharmonicities~\cite{peng,peano,melezhik}.

For the molecular branch, it has been shown 
theoretically
that the strictly one-dimensional
energies for the system without tunneling agree quite well 
with the full three-dimensional energies provided the
one-dimensional scattering length $a_{1\text{D}}$ is larger than the harmonic
oscillator length $a_{\rho}$~\cite{Gharashi14}.
Correspondingly, we restrict our
molecular branch calculations to this regime
(i.e., the smallest $a_{1\text{D}}$ considered in 
Sec.~\ref{sec_mol-two}---calculated 
using 
$\omega_{\rho} = \sqrt{p(t=0)} \omega_{\rho}^{\text{ref}}$---is 
$a_{1\text{D}} = 1.113 a_{\text{ho}}$, 
corresponding to $g_{1\text{D}} = -1.797 a_{\text{ho}}$). 
It should be kept in mind, however, that the validity regime of the 
one-dimensional framework could be different for static (energies)
and dynamic (tunneling) observables.

We use two different model interaction potentials $V_{\text{int}}$,
a zero-range potential $V_{\text{ZR}}$,
Eq.~(\ref{eq_VZR}),
and a finite-range gaussian potential $V_{\text{FR}}$,
\begin{eqnarray}
V_{\text{FR}}(z_{12}) = 
- V_{\text{G}} \exp \left(-  \frac{z_{12}^2}{2 z_0^2} \right),
\label{eq_VFR}
\end{eqnarray}
where $V_{\text{G}}$ and $z_0$ denote the depth ($V_{\text{G}} > 0$) and the range of the 
interaction.
We use $z_0 = 0.3 a_{\text{ho}}, 0.2 a_{\text{ho}}$ and $0.1 a_{\text{ho}}$, 
and adjust $V_{\text{G}}$ for each $z_0$ such that $V_{\text{FR}}$ yields the
desired one-dimensional two-body coupling constant $g_{1\text{D}}$.
Throughout, $V_{\text{G}}$ is chosen such that $V_{\text{FR}}$ supports at most
one even parity bound state in free space.
We find that the dependence of the tunneling observables on the range
$z_0$ is small.
This together with the fact that $V_{\text{ZR}}$ and $V_{\text{FR}}$
yield compatible tunneling results 
(as discussed below, we checked this for selected parameter combinations)
justifies the use of comparatively large $z_0$.
The real time propagation of the two-particle system
is discussed in Appendices~\ref{app_Cheb} and \ref{app_ZR-time}.

\begin{figure}[htbp]
\centering
\includegraphics[angle=0,width=0.4\textwidth, clip=true]{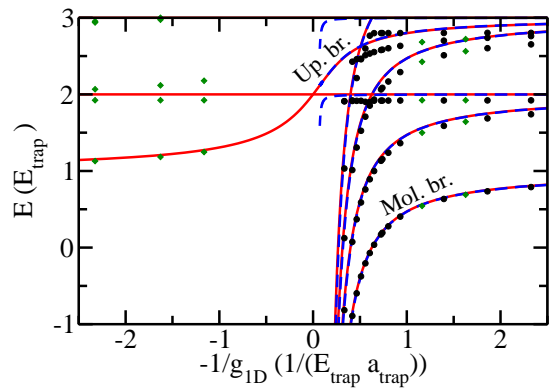}
\caption{(Color online)
Energies of two interacting trapped 
particles as a function of 
$-1/g_{1\text{D}}$.
Solid and dashed lines show the
energies for two particles with zero-range 
interaction 
and finite-range
interaction, respectively,
in a harmonic trap with frequency $\omega_{\text{trap}}$.
Circles and diamonds show the energies for two particles with
zero-range 
interaction
and finite-range interaction, respectively, in an 
anharmonic trap [see Eq.~(\ref{Vtrap})].
Both particles feel the same external potential 
[$p=0.795$, $c_{| j \rangle}(B) = 1.00115$, and $B' = 1892$G/m, corresponding 
to $\omega_{\text{trap}} = 2 \pi \times 1067.87$Hz].
The width of the finite-range potential is $z_0 = 0.0930 a_{\text{trap}}$.
}
\label{fig_energy}
\end{figure}

To get a first feeling for the two-particle system,
we consider the system with $p = p(-t_r)$ and map out the energy spectrum
as a function of $g_{1\text{D}}$.
We use the imaginary time propagation (see Appendix~\ref{app_im-time})
to find the ``eigenenergies'' and ``eigenfunctions'' of the system
[strictly speaking, the states are metastable due to the finite barrier 
for $p(-t_r) = 0.795$].
Figure~\ref{fig_energy}
shows the spectrum for two interacting particles described by the
Hamiltonian $H$, Eq.~(\ref{eq_H-2b}), with $p=0.795$ and 
$z_{\text{R}} = 8.548 a_{\text{ho}}$ as a function of $-1/g_{1\text{D}}$.
Both particles are assumed to feel the same single-particle trapping 
potential with $\mathcal{C} = 1894.18$G/m.
Diamonds and circles show the energies for the zero-range potential and
the finite-range potential with $z_0 = 0.1 a_{\text{ho}} = 0.0930 a_{\text{trap}}$,
respectively.
Note, throughout we use the zero-range potential to describe the positive 
$g_{1\text{D}}$ portion of the upper branch. In this regime,
the Hamiltonian with 
finite-range interaction supports many deep-lying states, making it 
challenging to select the low-energy states of interest
(recall, the relative and center-of-mass degrees of freedom are coupled).
Alternatively, one might consider
using a purely repulsive finite-range two-body potential. In this case,
however, a large $g_{1\text{D}}$ would require
a large range, thereby making the calculations model-dependent.
Hence, this alternative approach is not pursued here.
Figure~\ref{fig_energy}
uses the natural units $a_{\text{trap}}$ and $E_{\text{trap}}$ with 
$\omega_{\text{trap}} = 2 \pi \times 1067.87$Hz [see Eq.~(\ref{om-trap})].
The agreement between the zero-range and finite-range energies is very good 
for the $g_{1\text{D}}$ considered.

To illustrate the effect of the trap anharmonicity, solid and dashed lines
show the eigenspectrum for two particles interacting through 
$V_{\text{ZR}}$ and $V_{\text{FR}}$ under external harmonic confinement with 
frequency $\omega_{\text{trap}}$ (i.e., without 
magnetic 
field gradient and 
without anharmonicity).
The solid and dashed lines agree very well for most $g_{1\text{D}}$.
Differences are visible for the ``diving'' states near $1 / g_{1\text{D}} \approx 0$.
The differences arise because the states with odd relative parity
are not affected by the zero-range potential but are affected by the  
finite-range potential.
Comparing the energy spectrum for the isotropic trap (lines)
and the anharmonic trap (symbols), we see that the energies of the lowest
state agree well for negative $g_{1\text{D}}$ (molecular branch) and positive $g_{1\text{D}}$
(upper branch).
The negative $g_{1\text{D}}$ portion of the upper branch is affected comparatively 
strongly by the anharmonicity.
In this regime, the anharmonicity leads to a decrease of the energies due 
to the widening of the trap.
The coupling between the relative and center-of-mass degrees of freedom 
leads to avoided crossings between the energy levels that correspond,
for the harmonic trap, to even relative and odd relative parity states.
The eigenstates corresponding to the symbols on the upper branch and 
molecular branches serve as initial states for the real time evolution,
i.e., these states serve as our initial wave packets at $t = -t_r$.

\begin{figure}[htbp]
\centering
\includegraphics[angle=0,width=0.6\textwidth, clip=true]{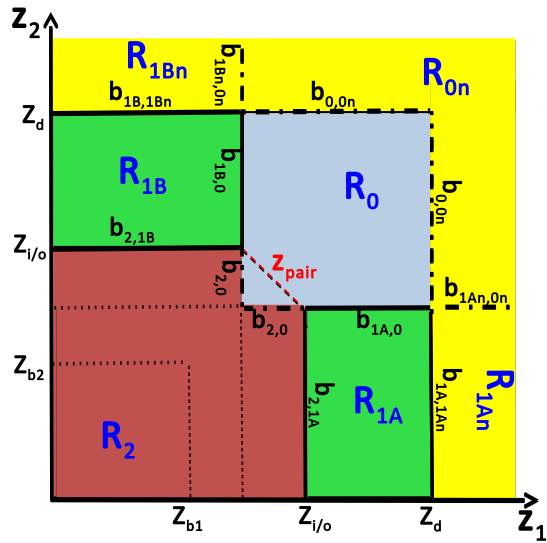}
\caption{(Color online)
Configuration space of 
the
two-particle system.
The regions $R_{j}$ ($j = 2, 1A, 1B, 0, 1An, 1Bn, 0n$) are shown in different 
colors/shades.
Each region $R_{j}$ is surrounded by the boundary $B_{j}$ (not shown).
Boundary segments that divide regions $R_{j}$ and $R_{j'}$ are labeled
by $b_{j,j'}$.
$z_{b1}$ and $z_{b2}$ denote the position of the maximum of the barrier
at $t = -t_r$.
$z_{\text{i}/\text{o}}$ divides the ``inside'' from the ``outside'';
we choose $z_{\text{i}/\text{o}}$ to be larger than $z_{b1}$ and $z_{b2}$
to ensure that the calculated flux is independent of how it is extracted.
$z_{\text{d}}$ denotes the largest $z_{1}$ and $z_{2}$ for which we calculate 
the ``physical'' wave packet.
$z_{\text{pair}}$ is equal to $2 a_{\text{1D}}$ for 
the
molecular branch and 
equal to
$2 a_{\text{ho}}$ for the upper branch;
$z_{\text{pair}}$ enters into our analysis of the pair tunneling
(see text for details).
}
\label{fig_regions}
\end{figure}

To analyze the tunneling dynamics of the two-particle system, we partition
the configuration space as shown in Fig.~\ref{fig_regions}.
Region $R_2$ corresponds to the situation where two particles are in the
trap, region $R_0$ corresponds to the situation where both particles have
left the trap, and region $R_{1A}$ ($R_{1B}$) corresponds to the situation 
where particle 1 (2) has left the trap while particle 2 (1) is in the trap.
The regions $R_{1An}$, $R_{1Bn}$, and $R_{0n}$ correspond to numerical regions
in which we apply damping
(see below). 
The region $R_j$ is encircled by the boundary $B_j$
(the $B_j$'s are not labeled in Fig.~\ref{fig_regions}).
To analyze the flux, the boundaries $B_j$ are broken up into boundary segments
$b_{j,j'}$ that border regions $R_j$ and $R_{j'}$.

The flux through boundaries $b_{2,1A}$ and $b_{2,1B}$ is
interpreted as uncorrelated single-particle tunneling while the flux 
through boundary $b_{2,0}$ is interpreted as pair tunneling.
The pair tunneling rate extracted from the flux through $b_{2,0}$ is not unique
and depends on $z_{\text{pair}}$.
Section~\ref{sec_mol-two} considers 
$1.1 a_{\text{ho}} < a_{\text{1D}} < 4.5 a_{\text{ho}}$;
motivated by the fact that the size of the free space molecule is 
approximately $a_{1\text{D}}$~\cite{footnote6},
we use $z_{\text{pair}} = 2 a_{1\text{D}}$
for this $a_{1\text{D}}$ range.
For the upper branch simulations, 
we use $z_{\text{pair}} = 2 a_{\text{ho}}$.
We found that the flux through $b_{2,0}$ is 
vanishingly small
for the upper branch simulations.
We set $z_{\text{i/o}}$, which defines where the ``inside'' region ends and the 
``outside'' region starts, 
such that $z_{\text{i/o}} > \max (z_{b1},z_{b2})$.
For the molecular branch simulations, the flux dynamics is
quite complex near the top of the barrier.
To be independent of the ``near-field'' dynamics,
we choose $z_{\text{i/o}} \approx 15 a_{\text{ho}}$ and
$13 a_{\text{ho}}$ for the molecular branch
and upper branch, respectively.
The physical regions end at $z_{\text{d}}$, i.e., for $z_1 > z_{\text{d}}$ or 
$z_2 > z_{\text{d}}$ a damping function is applied.
The damping function acts like an absorbing boundary
(see Appendix~\ref{app_damping}).
The damping function is needed since the flux reaches the end of the
simulation box within a small fraction of the total simulation time.
$z_{\text{d}}$ has to be so large that the two particles are essentially 
uncorrelated for $z > z_{\text{d}}$.
In practice we vary $z_{\text{d}}$ and choose its
value such that the observables do not change as $z_{\text{d}}$ is increased.
Typical values for $z_{\text{d}}$ are 
$25 a_{\text{ho}}$ for the molecular branch simulations and
$13 a_{\text{ho}}$ for the upper branch
simulations 
(for the upper branch, we found that
$z_{\text{d}}=z_{\text{i/o}}$
yields the same results as 
$z_{\text{d}}>z_{\text{i/o}}$).
As mentioned above, the time-dependent simulation starts at $t = - t_r$, 
where the probability $P_2(-t_r)$ to find two particles in the trap 
(i.e., in region $R_2$) equals 1.
For $t>t_r$, $P_2(t)$ decays with time. This decay, except for a short
period of time ($t \lesssim 20$ms), is well described by the exponential 
function
\begin{eqnarray}
P_2(t) =
P_2(t_{\text{ref}}) \exp[-\gamma_2 (t-t_{\text{ref}})],
\label{eq_gamma2}
\end{eqnarray}
where $\gamma_2$ denotes the decay rate.
Since both uncorrelated single-particle tunneling and pair tunneling can
contribute to the change of $P_2(t)$, we break $\gamma_2$ into two pieces,
$\gamma_2 = \gamma_{\text{s}} + \gamma_{\text{P}}$, where $\gamma_{\text{s}}$ 
and $\gamma_{\text{P}}$ denote the single-particle tunneling and pair tunneling
contributions, respectively (see Appendix~\ref{app_flux-anal} for details).
A non-zero $\gamma_{\text{s}}$ means that the probability $P_1(t)$ to find one 
particle in the trap is finite.
We 
also define the mean number $\bar N$ of trapped particles,
\begin{eqnarray}
\bar N(t) = 2 P_2(t) + P_1(t).
\label{eq_Nbar}
\end{eqnarray}
The time dependence of $\bar N(t)$ 
is
approximately parameterized by an
exponential decay with tunneling rate $\gamma$,
\begin{eqnarray}
\bar N(t) =
\bar N(t_{\text{ref}}) \exp[-\gamma (t-t_{\text{ref}})]
+ C,
\label{eq_Nbar-exp}
\end{eqnarray}
where $C$ denotes a constant.
Subsections~\ref{sec_mol-two} and
\ref{sec_upper-two} present the results of our time-dependent
two-particle simulations.

\section{Molecular branch tunneling dynamics}
\label{sec_molecular}

\subsection{Single-particle tunneling dynamics and trap calibration}
\label{sec_mol-single}
In the following we perform exact numerical calculations for the 
trap parameters reported in Table~\ref{table-exp}.
We will show that the numerically obtained tunneling rates do not agree 
with the measured ones and propose an alternative calibration approach.

\begin{figure}[htbp]
\centering
\includegraphics[angle=0,width=.35\textwidth, clip=true]{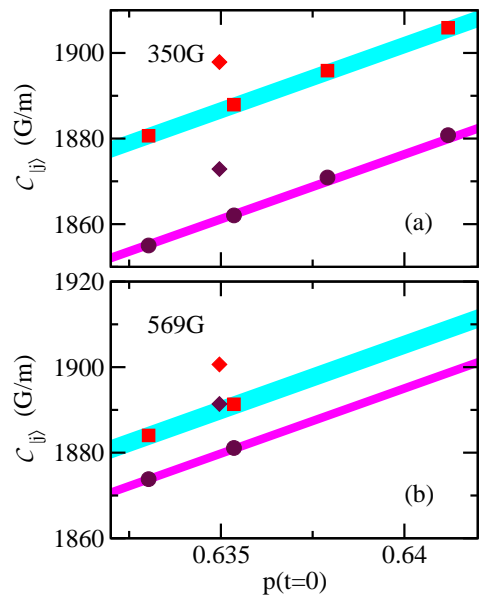}
\caption{(Color online) 
Parameter combinations ($p(t=0), \mathcal{C}_{|j \rangle} (B)$) 
that reproduce the 
experimentally measured single-particle tunneling rates at (a) $B=350$G
and (b)~$569$G.
For all calculations, $z_{\text{R}} = 9.975 \mu$m is used.
In panels (a) and (b), the initial state corresponds to the trap ground state.
The bands  
show the parameter combinations for which our full time-dependent calculations
reproduce the experimentally measured tunneling rates.
The widths of the bands originate from the experimental error
bars~\cite{Jochim4}.
In panels (a) and (b), the dark (magenta) and light (cyan) bands correspond to 
$^6$Li atoms in states $|1 \rangle$ and $|3 \rangle$, respectively.
Circles and squares show parameter combinations
for states $|1 \rangle$
and $|3 \rangle$, respectively, that are used in 
the two-particle calculations
(see Sec.~\ref{sec_mol-two}).
For comparison, the diamonds show the ($p(t=0)$, $\mathcal{C}_{|j \rangle}(B)$)
pairs that were suggested in Ref.~\cite{Jochim4}.
}
\label{fig_sptr-agreement}
\end{figure}

\begin{figure}[htbp]
\centering
\includegraphics[angle=0,width=0.4\textwidth, clip=true]{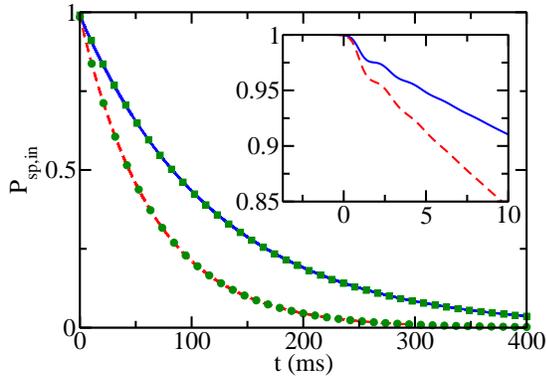}
\caption{(Color online)
Single-particle tunneling as a function of time for a $^6$Li atom in 
state $|1 \rangle$ at $B=350$G in the 
trap ground state.
The dashed and solid lines show the probability $P_{\text{sp,in}}(t)$ of finding 
the particle in the trap calculated using the exact time evolution for
$p(t=0) = 0.63496$ and $\mathcal{C}_{|1 \rangle}=1872.87$G/m 
($c_{|1 \rangle} = 0.98989$ and $B'=1892$G/m) 
(these parameters are proposed in Ref.~\cite{Jochim4}) 
and for $p(t=0) = 0.63536$ and $\mathcal{C}_{|1 \rangle}=1862.03$G/m
(this is one of many parameter sets that reproduces the experimentally 
measured tunneling rate), respectively.
The time evolution starts at $-t_r$ ($t_r \approx 3.72\text{ms}$ and 
$t_r \approx 3.71\text{ms}$ for the dashed and solid lines, respectively).
Circles and squares show exponentially decaying functions 
[see Eq.~(\ref{eq_exp-decay})] with $\gamma_{\text{sp}} = 15.39\text{s}^{-1}$ 
and 
$\gamma_{\text{sp}} = 8.28\text{s}^{-1}$, 
respectively.
The inset shows a blow-up of the short-time behavior.
}
\label{fig_sp-Pin}
\end{figure}

\begin{figure}[htbp]
\centering
\includegraphics[angle=0,width=.4\textwidth, clip=true]{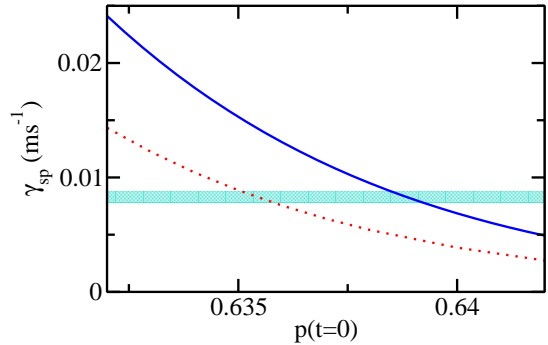}
\caption{(Color online)
Tunneling rate
of a  $^6$Li atom at $B=350$G
as a function of the dimensionless parameter $p(t=0)$.
The atom is prepared in the ground state of the trap, 
and $c_{ | 1 \rangle} = 0.98989$ and $B'=1892$G/m
are used. 
The solid and dotted lines show the tunneling rates obtained through exact
time propagation and the WKB approximation, respectively.
The horizontal band shows the tunneling rate 
$\gamma_{\text{sp}}^{\text{exp}} = 8.28(0.49) \text{s}^{-1}$
measured experimentally~\cite{Jochim4}
(the width of the band represents the experimental error bar).
}
\label{fig_sptr-wkb-num}
\end{figure}

The trap employed in the molecular branch experiments was 
calibrated, in addition to the calibration experiments already discussed in 
Sec.~\ref{sec_system}, based
on four single-particle experiments~\cite{Jochim4} 
[see Table~\ref{table-sp-1} 
and diamonds in Figs.~\ref{fig_sptr-agreement}(a) and
~\ref{fig_sptr-agreement}(b)].
In our first calculation, we use $\mathcal{C}_{|1 \rangle} = 1872.87$G/m,
corresponding to $c_{| 1 \rangle} = 0.98989$ and $B' = 1892$G/m,
and prepare the system in the trap ground state 
[see the diamond in Fig.~\ref{fig_sptr-agreement}(a)]. 
The dashed line in Fig.~\ref{fig_sp-Pin}
%
shows the result of our simulation for $p(0) = 0.63496$.
A fit of our data for $t>15$ms (the short-time dynamics exhibits, as can be 
seen in the inset of Fig.~\ref{fig_sp-Pin}, oscillations) to 
Eq.~(\ref{eq_exp-decay}) yields 
$\gamma_{\text{sp}}^{\text{num}} = 15.39 \text{s}^{-1}$
(see circles in Fig.~\ref{fig_sp-Pin}).
The tunneling rate 
$\gamma_{\text{sp}}^{\text{num}}$ obtained from the real time propagation
is 
nearly
twice as large as the experimentally measured tunneling rate 
$\gamma_{\text{sp}}^{\text{exp}}$, 
$\gamma_{\text{sp}}^{\text{exp}} = 8.28(0.49) \text{s}^{-1}$~\cite{Jochim4}.
This means that the trap parameters reported in Ref.~\cite{Jochim4},
obtained through the WKB analysis, yield a tunneling rate that deviates
by a factor of  
nearly
$2$ from the experimentally measured tunneling rate,
$\gamma_{\text{sp}}^{\text{num}} / \gamma_{\text{sp}}^{\text{exp}}=1.86$.
To understand this, we treat $t_r$ as a parameter.
Our tunneling simulations indicate that the exact shape of the initial state,
and thus $p(-t_r) V_0$, has a very small effect on the tunneling rate.
The tunneling rate, in contrast, 
depends appreciably on the value of $p(0) V_0$.
Thus, changing $t_r$ while keeping $p(-t_r) V_0$ fixed at $0.795 V_0$ has
a similar effect to changing $V_0$.
Solid and dotted lines in Fig.~\ref{fig_sptr-wkb-num}
%
%
show the numerically determined tunneling rate $\gamma_{\text{sp}}^{\text{num}}$ 
and the WKB tunneling rate $\gamma_{\text{sp}}^{\text{WKB}}$ as a function of 
$p(t=0)$, i.e., for varying $t_r$ 
(using $c_{| 1 \rangle}=0.98989$ and $B'=1892$G/m~\cite{Jochim4}).
It can be seen that the WKB analysis yields tunneling rates that differ by 
a factor of about 1/2 from those obtained from the full time evolution. 
This is elaborated on further in Appendix~\ref{appendix_wkb}.
Since the trap parameters reported by the experimental group utilized
the WKB approximation, we conclude that the trap parameters reported
in Table~\ref{table-exp} are inaccurate.
Table~\ref{table-sp-1}
%
\begin{table*}
\centering
\caption{
Experimentally measured single-particle tunneling rates 
$\gamma_{\text{sp}}^{\text{exp}}$
for selected 
magnetic field strengths
and initial single-particle 
states relevant to
the
molecular branch experiments~\cite{Jochim4}. 
Column 4 reports the values of
the dimensionless coefficients $c_{| j \rangle}(B)$
reported in 
Ref.~\cite{Jochim4}.
The fifth column reports the tunneling rate $\gamma_{\text{sp}}^{\text{num}}$ 
obtained from the exact time evolution using the trap parameters listed in 
Table~\ref{table-exp}.
As shown in column 6, the exact time evolution yields tunneling rates 
that are inconsistent with $\gamma_{\text{sp}}^{\text{exp}}$, suggesting 
that the trap calibration that involves the WKB analysis needs to be refined.
}
\begin{ruledtabular}
\begin{tabular}{l c c c c c} 
 state $| j \rangle $&
 $B$ (G) & 
 $\gamma_{\text{sp}}^{\text{exp}}$ ($\text{s}^{-1}$) & 
 $c_{| j \rangle}(B)$ & 
 $\gamma_{\text{sp}}^{\text{num}}$ ($\text{s}^{-1}$) & 
 $\gamma_{\text{sp}}^{\text{num}}/\gamma_{\text{sp}}^{\text{exp}}$
 \\
 \hline
 $| 1 \rangle $ trap's gr. st.&
 $350$  & 
 $8.28(0.49)$ &
 ~$0.98989$~ & 
 $15.39$ & 
 $1.86$ 
 \\  
 $| 3 \rangle $ trap's gr. st. & 
 $350$  & 
 $30.12(2.81)$ &
 ~$1.00311$~ & 
 $50.24$ &
 $1.67$
 \\
 $| 1 \rangle $ trap's gr. st.&
 $569$  & 
 $21.76(1.12)$ &
 ~$0.99968$~ & 
 $37.36$ & 
 $1.72$ 
 \\  
 $| 3 \rangle $ trap's gr. st. & 
 $569$  & 
 $35.25(3.57)$ &
 ~$1.00457$~ & 
 $55.87$ &
 $1.58$
 \\  
\end{tabular}
\end{ruledtabular}
\label{table-sp-1}
\end{table*}
%
compares the measured tunneling rates 
$\gamma_{\text{sp}}^{\text{exp}}$ with the numerically calculated tunneling 
rates 
$\gamma_{\text{sp}}^{\text{num}}$ for the trap
parameters summarized in Table~\ref{table-exp} and the 
$c_{| j \rangle}$ coefficients listed in Table~\ref{table-sp-1}.

The bands in Figs.~\ref{fig_sptr-agreement}(a) and~\ref{fig_sptr-agreement}(b)
show the $(p(t=0), \mathcal{C}_{|j \rangle})$ values for state $|1 \rangle$
[darker (magenta) band] and state $|3 \rangle$ [lighter (cyan) band] 
for which the
$\gamma_{\text{sp}}^{\text{num}}$ agree with the experimentally measured 
single-particle tunneling rates for states $|1 \rangle$ and $|3 \rangle$.
In our calculations, the initial state corresponds to the lowest trap state.
In a first attempt, we
did set $c_{|j \rangle}(B) = c_{|j \rangle}^{\text{BR}}(B)$
and aimed to find unique values for $p(t=0)$ and $B'$ that would reproduce 
all four experimentally measured tunneling rates.
For the functional form of the potential (with the parameters $V_0$,
$z_{\text{R}}$, $B'$, and $dp/dt$ from Table~\ref{table-exp}),
such a parameter combination does not exist.
Allowing $z_{\text{R}}$ to vary does not change the situation.
To reproduce the experimentally measured tunneling rates, we thus decided 
to treat $\mathcal{C}_{|j \rangle}(B)$ as a free parameter.
For example, we set 
$c_{|3 \rangle}(569\text{G}) = c_{|3 \rangle}^{\text{BR}}(569\text{G})$ 
and
$B'=1890$G/m and determine
$p(t=0)$ such that we reproduce the experimental single-particle rate.
We find $p(t=0) = 0.63536$.
We then set $p(t=0)$ to $0.63536$ and find $c_{|1 \rangle}(569\text{G})$, 
$c_{|1 \rangle}(350\text{G})$, and $c_{|3 \rangle}(350\text{G})$ 
such that $\gamma_{\text{sp}}^{\text{num}} = \gamma_{\text{sp}}^{\text{exp}}$
[see squares and circles in Figs.~\ref{fig_sptr-agreement}(a) 
and~\ref{fig_sptr-agreement}(b)].
We emphasize that 
these
are not unique parameter combinations.
Alternative parameter combinations that are also used in
Sec.~\ref{sec_mol-two} are marked in Figs.~\ref{fig_sptr-agreement}(a) 
and~\ref{fig_sptr-agreement}(b).

To obtain the $\mathcal{C}_{|j \rangle}(B)$ coefficients for other magnetic
fields, we use interpolations/extrapolations.
For state $|1 \rangle$, we use
\begin{eqnarray}
c_{|1 \rangle}(B) \approx c_0 + \frac{c_{-1}}{B} + \frac{c_{-2}}{B^2}
\label{eq_BR-expand}
\end{eqnarray} 
with $c_0=1.00338$, $c_{-1}=-1.89121$G, and $c_{-2} = -1565.12 \text{G}^2$.
This functional form (i) reproduces $c_{|1 \rangle}(350 \text{G}) = 0.985202$
and $c_{|1 \rangle}(569 \text{G}) = 0.995224$ and (ii) is designed such that 
the functional dependence of $c_{|1 \rangle}(B)$ is similar to that of
$c_{|1 \rangle}^{\text{BR}}(B)$.
For state $|3 \rangle$, we use $c_{|3 \rangle}(B) = c_{|3 \rangle}(569\text{G})$
for $B \ge 569$G and a linear interpolation for $350 \text{G} \le B \le 569$G
using the known $c_{|3 \rangle}$ values at 350G and 569G.
Table~\ref{table-used-param-molecular-1}
summarizes the parameters that are used in Sec.~\ref{sec_mol-two} 
to model the two-particle experiments.

\subsection{Two-particle tunneling dynamics}
\label{sec_mol-two}

This section considers two attractively-interacting
$^6$Li atoms in hyperfine states $|1\rangle$ and $|3\rangle$ 
on the molecular branch.
As discussed in Sec.~\ref{subsec_two-body_Ham},
the one-dimensional coupling constant $g_{1\text{D}}$
depends on $p(t)$. Specifically, $g_{1{\text{D}}}$
changes for
$t=-t_r$ to $t=0$ and is constant for $t=0$ to $t=t_{\text{hold}}$.
While this time dependence can be incorporated 
straightforwardly into the finite-range simulations
(in this case, the depth $V_{\text{G}}$ can be made to
vary with time), incorporating the time dependence into
the zero-range calculations is more involved since
$g_{1\text{D}}$ enters into the propagator.
To estimate the importance of the time dependence during the
initial down ramp (time $t=-t_r$ to $0$),
we compared the simulation results for the cases
where the full time dependence of $g_{1\text{D}}$ was
accounted for [i.e., $\omega_{\rho}$ was calculated
according to $\sqrt{p(t)} \omega_{\rho}^{\text{ref}}$]
and where the time dependence was neglected 
[i.e., $\omega_{\rho}$ was calculated
according to $\sqrt{p(0)} \omega_{\rho}^{\text{ref}}$]
for selected magnetic field strengths. We found that the difference 
between the resulting tunneling rates is between $0.02$\% and $0.2$\%.
Since this difference is much smaller than the difference
between our calculated 
tunneling rates and the experimentally
measured tunneling rates (see below),
the time dependence of $g_{1{\text{D}}}$ is neglected in 
what follows.
The reason why the tunneling rates, calculated
by accounting for and neglecting the time dependence of
$g_{1\text{D}}$, are so similar is two-fold.
First, very little tunneling occurs during the down ramp.
Second, the overlap between the states at $t=-t_r$ with somewhat
different $g_{1\text{D}}$ is much larger than the overlap between
the states at $t=-t_r$ and $t=0$. This implies that the down ramp
has a much larger effect on the state that results at $t=0$
than a small variation of $g_{1\text{D}}$ during the down ramp.

\begin{figure}[htbp]
\includegraphics[angle=0, width=0.45\textwidth, trim={6cm 0.05cm 6cm 0.0cm}, clip=true]{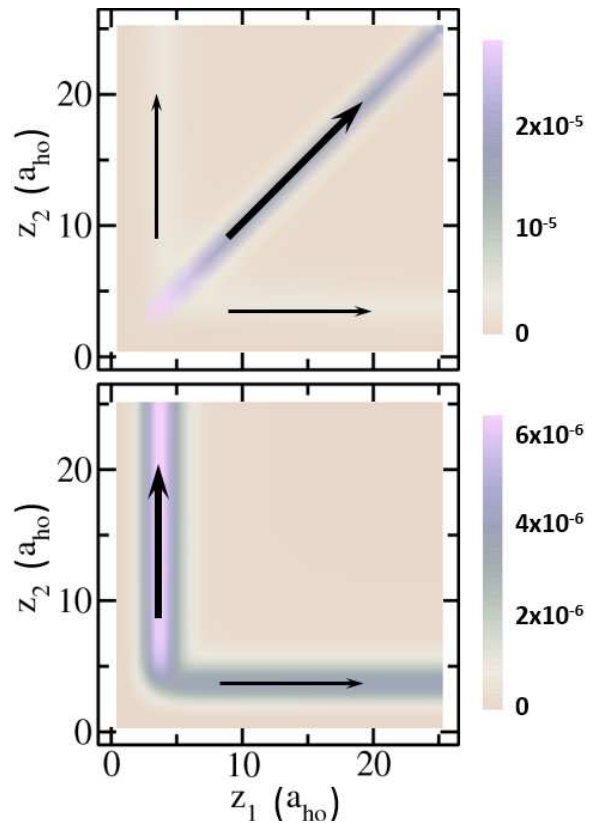}
\vspace*{0.1in}
\caption{(Color online)
Probability flux $|{\bf{j}}(z_1,z_2,t)|$.
The top and bottom panels show the probability flux at 
$t = 98$ms for two distinguishable particles with
$g_{1\text{D}}=-1.451 E_{\text{ho}} a_{\text{ho}}$ 
and $g_{1\text{D}}=0$, respectively
(the trap parameters are given,
respectively, 
in the 
sixth
and 
first
rows of
Table~\ref{table-used-param-molecular-1}).
The values of the flux are shown in the legend on the right
in units of $\omega / a_{\text{ho}}$
(note the different scales for the top and bottom panels).
The arrows indicate the primary directions of the flux ${\bf{j}}$.
}
\label{fig_res-mol-flux}
\end{figure}

The top panel in Fig.~\ref{fig_res-mol-flux}
shows the magnitude $|{\bf{j}}(z_1,z_2,t)|$ of the flux for 
$g_{1\text{D}}=-1.451 E_{\text{ho}} a_{\text{ho}}$ 
[corresponding to $a_{1\text{D}} / a_{\text{ho}} = 
1.378$ 
($B=1202$G, see Table~\ref{table-used-param-molecular-1})]
at $t = 98$ms.
As can be seen 
(see also the arrows in the top panel of Fig.~\ref{fig_res-mol-flux}),
the flux density is maximal along $z_1 \approx z_2$.
Only a 
small
portion of the flux is directed along the $\hat{\bf{z}}_1$ or
$\hat{\bf{z}}_2$ directions.
This demonstrates 
that pair tunneling becomes dominant for sufficiently strong interactions.
For comparison, the bottom panel of Fig.~\ref{fig_res-mol-flux} shows the 
quantity $|{\bf{j}}(z_1,z_2,t)|$ for the same $t$ but $g_{1\text{D}}=0$.
In this case pair tunneling is absent.
A careful comparison of the flux in the $\hat{\bf{z}}_1$ and
$\hat{\bf{z}}_2$ directions 
shows that the flux along $\hat{\bf{z}}_2$ is notably larger,
reflecting the fact that the trap felt by particle 2 
(parameterized via $\mathcal{C}_{| 3 \rangle}$) is shallower
than the trap felt by particle 1
(parameterized via $\mathcal{C}_{| 1 \rangle}$).
We note that the flux has a very intricate structure in the vicinity of the barrier, especially in the upper panel,
that is not visible on the scale of Fig.~\ref{fig_res-mol-flux}.
Unlike the flux plots shown in Fig.~3 of Ref.~\cite{Lundmark15},
we do not observe 
``wave-like patterns'' 
overlaying the flux.
We speculate that these features are artifacts of the numerics 
of Ref.~\cite{Lundmark15}.

\begin{figure}[htbp]
\centering
\includegraphics[angle=0,width=0.37\textwidth, clip=true]{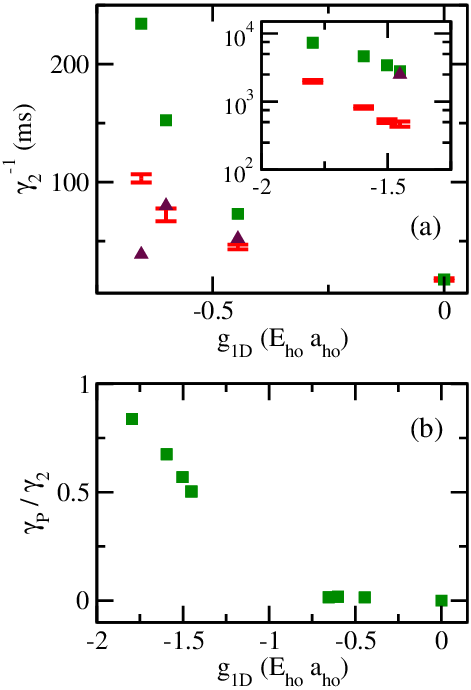}
\caption{(Color online)
Molecular branch tunneling dynamics for two distinguishable particles as 
a function of $g_{1\text{D}}$.
The coupling constant is calculated using 
$\omega_{\rho}=\sqrt{p(0)} \omega_{\rho}^{\text{ref}}$.
(a)~The squares show the results from our full time-dependent simulations
using the trap parameters given in Table~\ref{table-used-param-molecular-1};
these trap parameters yield single-particle tunneling rates
$\gamma_{\text{sp}}^{\text{num}}$ that agree with the experimentally measured 
single-particle tunneling rates $\gamma_{\text{sp}}^{\text{exp}}$.
The symbols with error bars show the experimental results~\cite{footnote5}.
For comparison, the triangles show the simulation results from
Ref.~\cite{Lundmark15}.
The inset shows the strongly-attractive region
using the same symbols as in the main figure
but a logarithmic 
$y$-scale.
(b)~Squares 
show the ratio $\gamma_{\text{P}} / \gamma_2$
obtained from our full time-dependent simulations.
}
\label{fig_res-mol-rates}
\end{figure}

Figure~\ref{fig_res-mol-rates} summarizes the tunneling rates obtained from our
full time-dependent molecular branch simulations
for finite-range interactions.
To obtain these results,
$\omega_{\rho}$ (and hence $g_{1{\text{D}}}$)
was calculated according to 
$\omega_{\rho} = \sqrt{p(t=0)} \omega_{\rho}^{\text{ref}}$.
Squares in 
Fig.~\ref{fig_res-mol-rates}(a) 
show the inverse of $\gamma_2$, i.e., the inverse of the rate with which the probability $P_2(t)$ to find both particles in the trap decays,
using the trap parameters that 
reproduce the experimentally measured single-particle tunneling rates.
As can be seen, the squares lie notably above the experimentally measured 
$(\gamma_2^{\text{exp}})^{-1}$ 
for finite $g_{1\text{D}}$;
for $g_{1\text{D}}=0$, the simulation results and the experimentally
measured rate agree by construction 
since the single-particle tunneling rates in this case add up to
$\gamma_2$ (see also Table~\ref{table-used-param-molecular-1}).

\begin{table*}
\caption{
Molecular branch dynamics for two distinguishable particles in states 
$| 1 \rangle$ and $| 3 \rangle$ for various magnetic 
field strengths.
The second column reports the one-dimensional coupling constant $g_{1\text{D}}$
calculated using $\omega_{\rho}=\sqrt{p(0)} \omega_{\rho}^{\text{ref}}$.
The third column indicates whether the simulation results were obtained using
the zero-range interaction model (ZR) or the gaussian interaction model with 
$z_0 = 0.2 a_{\text{ho}}$ (FR).
Columns 4 and 5 report the $\mathcal{C}_{| j \rangle}$ coefficients
for the trap parameterization 
with $p(t=0)=0.63536$ (see Sec.~\ref{sec_system})
and $z_{\text{R}} = 8.548 a_{\text{ho}}$.
Column 6 reports the tunneling rate $\gamma_2^{\text{num}}$ 
[see Eq.~(\ref{eq_gamma2})] 
obtained from our full time-dependent simulations. 
For comparison, column 7 shows the experimentally measured tunneling 
rates with error bars~\cite{footnote5}.
Column 8 shows the rate $\gamma_2^{\text{TI}}$ obtained from
time-independent 
simulations~\cite{Lundmark15}.
}
\begin{ruledtabular}
\begin{tabular}{c c c c c c c c } 
 $B$ (G) &  
 $g_{1\text{D}}$ $(a_{\text{ho}} E_{\text{ho}})$ &  
 ZR/FR &
 $\mathcal{C}_{| 1 \rangle}$ (G/m) & 
 $\mathcal{C}_{| 3 \rangle}$ (G/m)  &
 $\gamma_2^{\text{num}}$ ($\text{s}^{-1}$) &
 $\gamma_2^{\text{exp}}$ ($\text{s}^{-1}$)~\cite{footnote5} &
 $\gamma_2^{\text{TI}}$ ($\text{s}^{-1}$)~\cite{Lundmark15}
\\
  \hline
  \multirow{1}{*}{$569$} &
  $0$ &  
  ~---~ &
 ~$1881.11$ ~ & 
 ~$1891.32$~ & 
 ~$57.0$~ & 
 ~$57.01(3.74)$~  &
 ~---~
\\
 \multirow{1}{*}{$496$} &
  $-0.446$ &  
  ZR  &
 ~$1877.16$ ~ & 
 ~$1890.19$~ & 
 ~$13.8(0.3)$~ & 
 ~$22.2(1.0)$~ &
 ~$19.2(0.5)$~ 
\\
 \multirow{1}{*}{$496$} &
  $-0.446$ &  
  FR  &
 ~$1877.16$ ~ & 
 ~$1890.19$~ & 
 ~$14.0$~ & 
 ~$22.2(1.0)$~ &
 ~$19.2(0.5)$~ 
\\
 \multirow{1}{*}{$423$} &
  $-0.601$ &  
  FR  &
 ~$1871.41$ ~ & 
 ~$1889.06$~ & 
 ~$6.67$~ & 
 ~$13.84(1.04)$~  &
 ~$12.5(0.5)$~
\\
 \multirow{1}{*}{$350$} &
  $-0.654$ &  
  FR  &
 ~$1862.03$ ~ & 
 ~$1887.93$~ & 
 ~$4.27$~ & 
 ~$9.70(0.33)$~  &
 ~$25.8(0.5)$~
\\
 \multirow{1}{*}{$1202$} &
  $-1.451$ &  
  FR  &
 ~$1891.38$ ~ & 
 ~$1891.32$~ & 
 ~$0.360$~ & 
 ~$2.14(0.19)$~  &
 ~$0.4(0.5)$~
\\
 \multirow{1}{*}{$1074$} &
 $-1.503$ &  
  FR  &
 ~$1890.49$ ~ & 
 ~$1891.32$~ & 
 ~$0.293$~ & 
 ~$1.931(0.123)$~  &
 ~---~
\\
 \multirow{1}{*}{$958$} &
 $-1.595$ &  
  FR  &
 ~$1889.44$ ~ & 
 ~$1891.32$~ & 
 ~$0.216$~ & 
 ~$1.227(0.053)$~  &
 ~---~
\\
 \multirow{1}{*}{$851$} &
 $-1.797$ &  
  FR  &
 ~$1888.11$ ~ & 
 ~$1891.32$~ & 
 ~$0.137$~ & 
 ~$0.505(0.023)$~  &
 ~---~
\\
\end{tabular}
\end{ruledtabular}
\label{table-used-param-molecular-1}
\end{table*}

\begin{table}
\caption{
Molecular branch dynamics for two distinguishable particles in states 
$| 1 \rangle$ and $| 3 \rangle$ for various magnetic 
field strengths.
The second column reports the one-dimensional coupling constant $g_{1\text{D}}$
calculated using $\omega_{\rho}=0.67 \sqrt{p(0)} \omega_{\rho}^{\text{ref}}$.
The calculations are performed using the gaussian interaction model with 
$z_0 = 0.2 a_{\text{ho}}$ and the trap parameters are the same as
those for the
calculations reported in Table~\ref{table-used-param-molecular-1}.
Column 3 reports the tunneling rate $\gamma_2^{\text{num}}$ 
[see Eq.~(\ref{eq_gamma2})] 
obtained from our full time-dependent simulations. 
}
\begin{ruledtabular}
\begin{tabular}{l c c}
 $B$ (G) & $g_{1\text{D}}$ ($E_{\text{ho}}a_{\text{ho}}$) & 
$\gamma_2^{\text{num}}$ ($s^{-1}$) \\
\hline
 $496$ & $-0.303$ & $22.4$ \\
 $423$ & $-0.410$ & $13.4$ \\
 $350$ & $-0.447$ & $9.57$ \\
 $1202$ & $-1.018$ & $1.89$ \\
 $1074$ & $-1.056$ & $1.53$ \\
 $958$ & $-1.124$ & $1.04$ \\
 $851$ & $-1.275$ & $0.62$ \\
\end{tabular}
\end{ruledtabular}
\label{table_new}
\end{table}

The molecular branch tunneling dynamics has previously been calculated
by Lundmark {\it{et al.}} using a time-independent method~\cite{Lundmark15}.
Unfortunately, 
the trap parameters used to perform the 
calculations
were not reported. 
The triangles in 
Fig.~\ref{fig_res-mol-rates}(a) 
show the result
of this study. It can be seen that 
the inverse tunneling rate $(\gamma_2^{\text{num}})^{-1}$ 
is a 
non-monotonic
function of $g_{1\text{D}}$;
such non-monotonic behavior is not displayed in our simulations.
Reference~\cite{Lundmark15} 
interpreted
the
non-monotonic
dependence 
as an interplay between the trap
parameters.

To quantify the contribution of pair tunneling, we break $\gamma_2$ into two
parts, $\gamma_2 = \gamma_{\text{P}} + \gamma_{\text{s}}$, where 
$\gamma_{\text{P}}$ is the pair tunneling rate and $\gamma_{\text{s}}$ 
the single-particle tunneling rate.
We identify these rates from the flux passing through 
the boundary $b_{2,0}$ and the sum of the fluxes
passing through the boundaries $b_{2,1A}$ and $b_{2,1B}$. 
Figure~\ref{fig_res-mol-rates}(b) 
shows the ratio
$\gamma_{\text{P}} / \gamma_2$ as a function of the interaction strength.
We find
that
$\gamma_{\text{P}}$
is approximately equal to
$0$ for
$g_{1\text{D}} \ge -0.654 E_{\text{ho}} a_{\text{ho}}$.
As one might predict intuitively,
the ratio $\gamma_{\text{P}} / \gamma_2$ increases to close to 1 for 
stronger attractive interactions.
In this regime, the molecule can be treated as a point particle
of mass $2m$.
Our simulation results for 
$g_{1\text{D}} \ge -0.654 E_{\text{ho}} a_{\text{ho}}$ 
are consistent 
with the
experimental observation of negligibe pair tunneling. 
In the strongly-interacting regime, i.e., for
$g_{1\text{D}} \le -1.451 E_{\text{ho}} a_{\text{ho}}$, the experiments 
could not resolve the pair versus single-particle tunneling fractions.

To understand why
our finite-$g_{1\text{D}}$ simulations predict larger tunneling constants
$1/\gamma_2$ than measured experimentally
[see Fig.~\ref{fig_res-mol-rates}(a)],
we repeated our simulations using several possible 
parameter sets
that reproduce the experimentally measured single-particle tunneling rates,
marked on the bands in Fig.~\ref{fig_sptr-agreement}.
We found that the
two-body results remain almost unchanged, suggesting that the
non-uniqueness of the trap parameterization is not the cause 
for the disagreement.
We also repeated one calculation using  the
zero-range interaction model as opposed to the finite-range interaction model
(see Table~\ref{table-used-param-molecular-1}).
Again, we found that the result remains almost unchanged,
suggesting that finite-range effects 
are not the cause 
for the disagreement.
As a third possibility we investigated the dependence 
of the tunneling rates on $\omega_{\rho}$. As we now show,
a smaller $\omega_{\rho}$ brings the tunneling rates 
obtained from the full time-dependent simulations
in pretty good
agreement with the experimentally
measured tunneling rates.

\begin{figure}[htbp]
\centering
\includegraphics[angle=0,width=0.37\textwidth, clip=true]{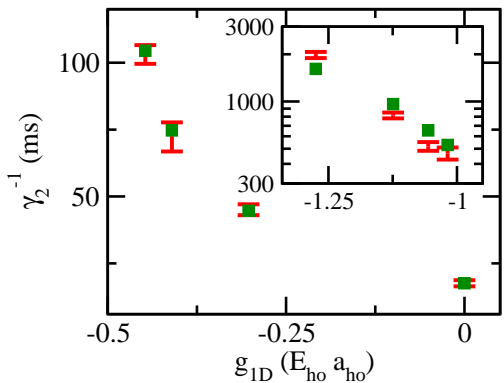}
\caption{(Color online)
Molecular branch tunneling dynamics for two distinguishable particles as 
a function of $g_{1\text{D}}$.
The coupling constant is calculated using 
$\omega_{\rho}=0.67 \sqrt{p(0)} \omega_{\rho}^{\text{ref}}$.
The squares show the results from our full time-dependent simulations
using the trap parameters given in Table~\ref{table-used-param-molecular-1};
these trap parameters yield single-particle tunneling rates
$\gamma_{\text{sp}}^{\text{num}}$ that agree with the experimentally measured 
single-particle tunneling rates $\gamma_{\text{sp}}^{\text{exp}}$.
The symbols with error bars show the experimental results~\cite{footnote5}.
The inset shows the strongly-attractive region
using the same symbols as in the main figure
but a logarithmic 
$y$-scale.
}
\label{fig_new}
\end{figure}

As discussed in Sec.~\ref{subsec_two-body_Ham}
the magnetic field gradient pushes the particles 
out to finite positive $z$,
resulting in, on average, a
weaker confinement along the tight
confinement direction.
Squares in Fig.~\ref{fig_new} show $1/\gamma_2$, obtained
from our full time-dependent simulations, using the trap parameters that
reproduce the experimentally measured single-particle tunneling rates
and $g_{1{\text{D}}}$ calculated according to 
$\omega_{\rho}= 0.67 \sqrt{p(t=0)} \omega_{\rho}^{\text{ref}}$
(see also Table~\ref{table_new}).
The factor of $0.67$ yields (roughly) maximal agreement between
the
time constants obtained from our simulations and 
those measured experimentally
(symbols with error bars in Fig.~\ref{fig_new}). 
Recalling the discussion
presented in Sec.~\ref{subsec_two-body_Ham}, this value
seems reasonable, though possibly slightly smaller than one might
have expected naively.
While other explanations for the disagreement between the
squares and the symbols with error bars in 
Fig.~\ref{fig_res-mol-rates}(a) cannot be ruled out,
our results indicate that the addition of the magnetic field
gradient may have a non-trivial effect on the 
calculation of
the renormalized one-dimensional coupling constant
$g_{1{\text{D}}}$.

\section{Upper branch tunneling dynamics}
\label{sec_upper}

\subsection{Trap calibration}
\label{sec_upper-trp-cal}

As 
discussed 
in Sec.~\ref{sec_system}, the trap used in the
upper branch experiment was calibrated by preparing two
identical non-interacting fermions in state $|2 \rangle$ 
at 
various magnetic field strengths.
The measured tunneling 
rates 
$\gamma^{\text{exp}}$ 
were
obtained
by fitting $\bar{N}(t)$ to an exponential 
plus a constant.
Table~\ref{table-used-param-upper}
%
\begin{table*}
\caption{
Upper branch dynamics for two distinguishable particles in states $| 1 \rangle$
and $| 2 \rangle$ for various magnetic field
strengths.
The second column reports the one-dimensional coupling constant $g_{1\text{D}}$
calculated using 
$\omega_{\rho}= \sqrt{p(0)} \omega_{\rho}^{\text{ref}}$.
The third column indicates whether the simulation results were obtained using
the zero-range interaction model (ZR) or the gaussian interaction model with 
$z_0 = 0.2 a_{\text{ho}}$ (FR).
Columns 4 and 5 report the $\mathcal{C}_{| j \rangle}$ coefficients; as 
discussed in Sec.~\ref{sec_upper-trp-cal}, we use
$c_{| j \rangle} = c_{| j \rangle}^{\text{BR}}$,
$B' = 1890$G/m,
$p(t=0)=0.68$, 
and $z_{\text{R}} = 8.548 a_{\text{ho}}$.
Column 6 reports the tunneling rate $\gamma^{\text{num}}$ 
[see Eq.~(\ref{eq_Nbar-exp})] 
obtained from our full time-dependent simulations. 
For comparison, column 7 shows the experimentally measured tunneling 
rates with 
error bars~\cite{footnote11}.
}
\begin{ruledtabular}
\begin{tabular}{c c c c c c c} 
  $B$ (G) &  
 $g_{1\text{D}}$ ($a_{\text{ho}} E_{\text{ho}}$) &
 ZR/FR &
 $\mathcal{C}_{| 1 \rangle}$ (G/m) & 
 $\mathcal{C}_{| 2 \rangle}$ (G/m) &
 $\gamma^{\text{num}}$ ($\text{s}^{-1}$) &
 $\gamma^{\text{exp}}$ ($\text{s}^{-1}$)
\\
 \hline
 $750$ &
 $6.15$ &
 ZR &
 $1883.86$  & 
 $1881.60$ &
 $4.2(0.5)$ &
 $2.9(0.2)$ 
\\
 $782$ &
 $\infty$ &
 ZR &
 $1884.56$  & 
 $1882.47$ &
 $15$ &
 $12.8(1.2)$ 
\\
 $855$ &
 $-4.42$ &
 FR &
 $1885.88$  & 
 $1884.07$ &
 $77$ &
 $62.8(8.2)$ 
\\
 \multirow{1}{*}{$900$} &
 $-3.15$ &
 ZR &
 $1886.57$  & 
 $1884.90$ &
 $127$ &
 $107(12)$ 
\\
 \multirow{1}{*}{$900$} &
 $-3.15$ &
 FR &
 $1886.57$  & 
 $1884.90$ &
 $130$ &
 $107(12)$ 
 \end{tabular}
\end{ruledtabular}
\label{table-used-param-upper}
\end{table*}
%
summarizes 
$\gamma^{\text{exp}}$~\cite{footnote11}.
To see if the trap parameterization proposed by the experimental group is
accurate, we perform
a time-dependent two-particle simulation for 
the
anti-symmetrized two-particle wave packet using the trap parameters
reported in Table~\ref{table-exp} and $c_{|2 \rangle}=1$.
We find $\gamma^{\text{num}}=6.86 \text{s}^{-1}$, which is about two times 
smaller than the experimentally measured value, i.e.,
$\gamma^{\text{num}} / \gamma^{\text{exp}} \approx 0.5$ 
(note, this ratio is around 1.7 for the molecular branch;
see Sec.~\ref{sec_mol-single} and
Appendix~\ref{appendix_wkb}).
Similar to the molecular branch, we conclude that the WKB approximation
cannot be used to calibrate the trap.

To recalibrate the trap, we set $c_{|2 \rangle} = c_{|2 \rangle}^{\text{BR}}$
and adjust $p(t=0)$ and $B'$ such that 
$\gamma^{\text{num}}$ for the anti-symmetric two-particle state
at $B=782$G agrees, within error bars,
with the experimentally measured tunneling rate.
As in the molecular branch (see Fig.~\ref{fig_sptr-agreement}),
we do not find a unique parameter combination
but  a parameter band. 
Using $p(t=0)=0.68$,
$B'=1890$G and $c_{|2\rangle} = c_{|2 \rangle}^{\text{BR}}$,
we find the tunneling rate $\gamma^{\text{num}}$ for several 
magnetic field strengths
(see 
Table~\ref{table-used-param-upper-NI}).
%
\begin{table}
\caption{
Tunneling dynamics for two identical particles in state 
$| 2 \rangle$ for various magnetic field
strengths.
The second column reports the $\mathcal{C}_{| j \rangle}$ coefficients; as 
discussed in Sec.~\ref{sec_upper-trp-cal}, we use
$c_{| j \rangle} = c_{| j \rangle}^{\text{BR}}$,
$B' = 1890$G/m,
$p(t=0)=0.68$, and $z_{\text{R}} = 8.548 a_{\text{ho}}$.
Column 3 reports the tunneling rate $\gamma^{\text{num}}$ 
[see Eq.~(\ref{eq_Nbar-exp})] 
obtained from our full time-dependent simulations. 
For comparison, column 4 shows the experimentally measured tunneling 
rate $\gamma^{\text{exp}}$ with 
error bars~\cite{footnote11}.
}
\begin{ruledtabular}
\begin{tabular}{c c c c} 
 $B$ (G) &  
 $\mathcal{C}_{| 2 \rangle}$ (G/m) &
 $\gamma^{\text{num}}$ ($\text{s}^{-1}$) &
 $\gamma^{\text{exp}}$ ($\text{s}^{-1}$)
\\
 \hline
 $750$ &
 $1881.60$ &
 $13.2$ &
 $14.7(1.3)$ 
\\
 $782$ & 
 $1882.47$ &
 $13.8$ &
 $13.2(1.1)$ 
\\
 $820$ &
 $1883.37$ &
 $14.5$ &
 $13.1(1.4)$ 
\\
 $855$ &
 $1884.07$ &
 $15.1$ &
 $11.5(1.3)$ 
\\
 \multirow{1}{*}{$900$} &
 $1884.90$ &
 $15.8$ &
 $16.0(1.1)$  
 \end{tabular}
\end{ruledtabular}
\label{table-used-param-upper-NI}
\end{table}
%
Our $\gamma^{\text{num}}$ agree with $\gamma^{\text{exp}}$
within error bars, except for the 
cases at $B=750$G and 
$B=855$G, where the deviations are, respectively, about 
$1.1$ and $2.5$
times larger than the error
bars.


\subsection{Two-particle tunneling dynamics}
\label{sec_upper-two}

\begin{figure}[htbp]
\centering
\includegraphics[angle=0,width=0.38\textwidth, clip=true]{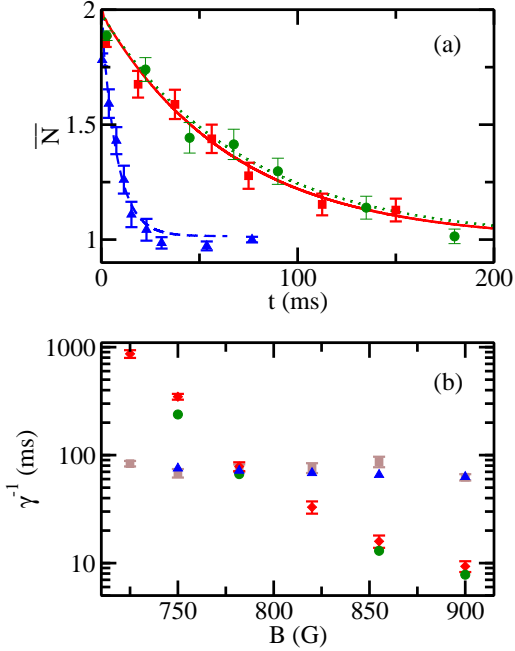}
\caption{(Color online) Upper branch tunneling dynamics. (a)~The 
dashed and solid lines show the mean number of trapped particles $\bar N$ obtained from our full time-dependent simulations as a function of time for two distinguishable particles at $B=900$G and $B=782$G, respectively, using the trap and interaction parameters given in Table~\ref{table-used-param-upper}. For comparison, triangles and squares with error bars show the corresponding 
experimental 
results~\cite{footnote11}. 
The dotted line shows the mean number of trapped particles $\bar N$ obtained 
from our full time-dependent simulations as a function of time for two identical fermions at $B=782$G, using the trap 
parameters given in Table~\ref{table-used-param-upper-NI}. 
For comparison, circles show the corresponding experimental results. (b)~Circles and triangles show the time constant $\gamma^{-1}$ obtained from our full 
time-dependent simulations for two distinguishable particles and two identical fermions, respectively, as a function of the magnetic 
field 
strength
$B$. For comparison, diamonds and squares with error bars show the corresponding experimental 
results~\cite{footnote11}.
}
\label{fig_res-rep-1}
\end{figure}

This section discusses the upper branch tunneling dynamics
for two distinguishable particles
with finite interaction strength 
$g_{1{\text{D}}}$.
Solid and dashed lines in Fig.~\ref{fig_res-rep-1}(a) show the mean number of
trapped particles $\bar N$, Eq.~(\ref{eq_Nbar}), extracted from our full 
time-dependent simulations as a function of the hold time for two 
distinguishable particles at $B=782$G 
($g_{\text{1D}} = 192 a_{\text{ho}} E_{\text{ho}}$;
in what follows, we use $g_{1{\text{D}}}=\infty$ for this magnetic field
strength) 
and $B=900$G 
($g_{\text{1D}} = -3.15 a_{\text{ho}} E_{\text{ho}}$).
Here, $g_{1{\text{D}}}$ is calculated using 
$\omega_{\rho} = \sqrt{p(t=0)} \omega_{\rho}^{\text{ref}}$.
As can be seen in Fig.~\ref{fig_energy}, the upper branch energy of the 
quasi-eigenstate at $t = -t_{\text{r}}$ is larger for negative $g_{\text{1D}}$
than for infinitely large $g_{\text{1D}}$.
This implies that the effective barrier height that the two-particle system 
sees is smaller at $B=900$G than at $B=782$G, resulting in faster tunneling 
dynamics for the system at $B=900$G than at $B=782$G.
The tunneling rates $\gamma$, obtained by fitting our data to 
Eq.~(\ref{eq_Nbar-exp})
or from the flux analysis 
(see Appendix~\ref{app_flux-anal}),
are $\gamma^{\text{num}} = 127 \text{s}^{-1}$ 
for $B=900$G and $\gamma^{\text{num}} = 15 \text{s}^{-1}$ 
for $B=782$G.
These tunneling rates agree 
at the two sigma level
with the experimentally measured rates
of $\gamma^{\text{exp}} = 107(12) \text{s}^{-1}$ 
[see triangles in Fig.~\ref{fig_res-rep-1}(a)]
and $12.8(1.2) \text{s}^{-1}$~\cite{footnote11} 
[see squares in Fig.~\ref{fig_res-rep-1}(a)].

An important aspect of the tunneling dynamics of the upper branch is that the 
mean number of trapped particles $\bar N$ decreases from 2 to 
approximately
1 over the hold 
times considered. 
This suggests that the particle that remains trapped has such a small
energy that its tunneling dynamics is orders of magnitude slower than the 
tunneling dynamics considered in Fig.~\ref{fig_res-rep-1}.
Indeed, we observe essentially no flux through the boundaries $b_{1A,0}$
and $b_{1B,0}$.
Comparing the portion of the wave packet in region $R_{1A}$ (or $R_{1B}$)
with the quasi-eigenstate of a single trapped particle shows that the remaining
particle occupies to a good approximation the lowest trap state.
This  implies that the particle that leaves the trap carries away the 
``excess energy''.
Performing single-particle calculations for particles $| 1 \rangle$ and
$| 2 \rangle$ initially in the trap ground state, we find tunneling rates of
0.008s$^{-1}$ and 0.007s$^{-1}$.
This confirms the separation of time scales alluded to above.

Circles in Fig.~\ref{fig_res-rep-1}(b) show our tunneling time constants
$(\gamma^{\text{num}})^{-1}$ for two distinguishable particles as a function 
of the magnetic field
strength.
Our $(\gamma^{\text{num}})^{-1}$ 
follow
the overall trend of the experimentally 
measured $(\gamma^{\text{exp}})^{-1}$ [diamonds in Fig.~\ref{fig_res-rep-1}(b)]
but lie a bit lower (see also Table~\ref{table-used-param-upper}). 
The discrepancy is largest for positive $g_{\text{1D}}$ ($B=750$G), where
the dynamics is slowest.
This is the regime where our simulations are, due to the slow tunneling,
the most demanding. 
We estimate, however, that our numerical uncertainties do not account for the
$45\%$ discrepancy 
between the calculated tunneling constant $(\gamma^{\text{num}})^{-1}$
and the experimentally measured tunneling constant $(\gamma^{\text{exp}})^{-1}$.

Motivated by the analysis presented in Sec.~\ref{sec_mol-two},
one may ask how the tunneling rates for the upper branch
depend on the $\omega_{\rho}$ value used to calulate $g_{1{\text{D}}}$.
We estimate that a scaling factor of around $0.85$
improves
the agreement between our simulations and the experiment
for $B=750$G; at the same time, the agreement for $B=855$G and $B=900$G
detoriates.
The fact that the ``optimal''
scaling factor for the upper branch seems to differ
from that for the molecular branch is not unreasonable.
First, since the non-linear trap term is larger, one might expect
that $\omega_{\rho}$ is modified less by the magnetic field
gradient term for the upper branch than 
for the molecular branch. 
Second, the excited upper branch
states may be affected differently
than the molecular branch states
[one should keep in mind that Eq.~(\ref{eq_olsh}) is
an approximation].

%
%

It is interesting to compare, as has been done in the experiments, the 
tunneling dynamics for two distinguishable particles with that for two 
identical particles,
since two distinguishable but otherwise identical particles with infinitely
large $g_{1\text{D}}$ are known to become 
fermionized~\cite{Olsh98,Girardeau03, Kanjilal04}.
In the present case, the distinguishable particles in states 
$|1\rangle$ and $|2\rangle$ feel slightly different trapping potentials. 
Thus the fermionization concept does, strictly speaking, not apply.
However, since $\mathcal{C}_{| 1 \rangle}$ and $\mathcal{C}_{| 2 \rangle}$
at $B=782$G differ by only $0.2\%$, a meaningful comparison can be made. The dotted line in Fig.~\ref{fig_res-rep-1}(a) shows the mean number of particles for two identical fermions in state $| 2 \rangle$. 
Since $\mathcal{C}_{| 2 \rangle}(782\text{G}) < \mathcal{C}_{| 1 \rangle}$(782G), implying a higher barrier for the atom in state $| 2 \rangle$ than the atom in state $| 1 \rangle$, the non-interacting identical fermion system (two atoms in state $| 2 \rangle$) tunnels slightly slower than the two distinguishable atom system (one atom in state $| 1 \rangle$ and one atom in state $| 2 \rangle$) with infinitely large $g_{1\text{D}}$.
Triangles and squares in Fig.~\ref{fig_res-rep-1}(b) show the tunneling 
constants 
$\gamma^{-1}$ for two identical fermions as a function of $B$ obtained from our 
simulations 
(see Sec.~\ref{sec_upper-trp-cal} and Table~\ref{table-used-param-upper-NI})
and from experiment, respectively.
Although the fermionization is only approximate, Fig.~\ref{fig_res-rep-1}(b) shows that the tunneling rate curves for two distinguishable particles and two identical non-interacting fermions cross at approximately $B=782$G, corresponding to $g_{1\text{D}} = \infty$ for the $| 1 \rangle$--$| 2 \rangle$ interaction.

\begin{figure}[htbp]
\centering
\includegraphics[angle=0,width=0.4\textwidth, clip=true]{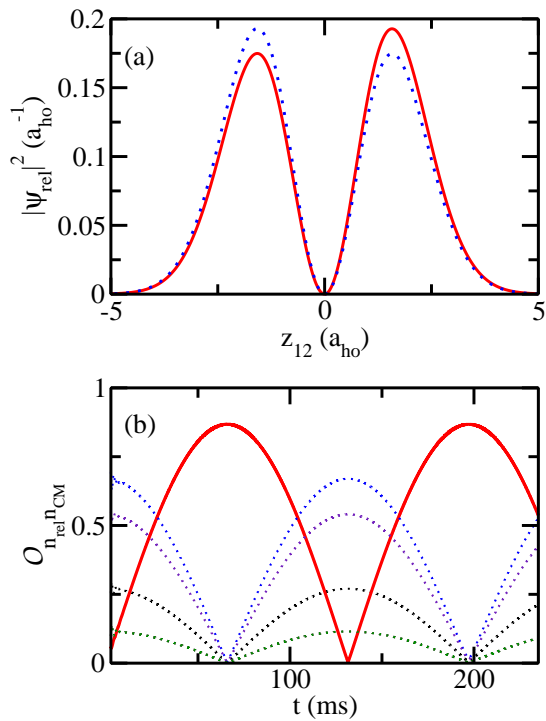}
\caption{(Color online)
Analysis of the upper branch time dynamics for two distinguishable particles
interacting through a zero-range potential with $g_{1\text{D}}=\infty$
(the trap parameters are given in row 2 of 
Table~\ref{table-used-param-upper}).
(a)~The solid and dotted lines show the density $|\Psi_{\text{rel}}(z_{12})|^2$
of the lowest ``eigenstate'' and the first excited ``eigenstate''
at $t=-t_r$.
These states are nearly degenerate.
(b)~The dotted lines show the normalized overlap $\mathcal{O}_{n_{\text{rel}}n_{\text{CM}}}$ 
between the wave packet $\Psi(z_1,z_2,t)$ and the harmonic oscillator states
with even relative parity 
[from top to bottom, $(n_{\text{rel}},n_{\text{CM}}) =$ 
(0,0), (2,0), (0,1), and (0,2)].
The solid line shows the overlap $\mathcal{O}_{n_{\text{rel}}n_{\text{CM}}}$ 
between the wave packet $\Psi(z_1,z_2,t)$ and the lowest harmonic oscillator 
state with odd relative parity, i.e., with $(n_{\text{rel}},n_{\text{CM}})=(1,0)$.
The harmonic oscillator states are characterized by $a_{\text{trap}} = 1.073 a_{\text{ho}}$.
}
\label{fig_res-rep-ginf}
\end{figure}

Another consequence of the fact that
$\mathcal{C}_{| 2 \rangle}(782\text{G}) < \mathcal{C}_{| 1 \rangle}$(782G)
is that the 
probability to find the particle ordering $z_1 < z_2$ (or $z_1 > z_2$)
for two atoms in states $| 1 \rangle$ and $| 2 \rangle$
changes as a function of time.
At $t = -t_r$, the probability $P_{z_1>z_2}$ to find $z_1 > z_2$ is 0.525 and 
the probability $P_{z_1<z_2}$ to find $z_1 < z_2$ is 0.475
[see Fig.~\ref{fig_res-rep-ginf}(a)].
This is due to the fact that 
the 
particle 
in state
$| 1 \rangle$ feels a ``softer'' 
confinement than the particle in state $| 2 \rangle$,
i.e., $\omega_{\text{trap}}$ for state $| 1 \rangle$ is less than 
$\omega_{\text{trap}}$ for state $| 2 \rangle$.
Importantly, the particles in states $| 1 \rangle$ and $| 2 \rangle$ at 
$B=782$G ($g_{\text{1D}} = \infty$) cannot pass through each other.
Thus, since the particle in state $| 1 \rangle$ tunnels slightly faster than 
the particle in state $| 2 \rangle$ (see below)
the probability $P_{z_1>z_2}$ to have $z_1 > z_2$ inside the trap gets depleted
faster than the probability to have $z_1 < z_2$.
Indeed, at $t = 94$ms, we have $P_{z_1<z_2} = P_{z_1>z_2}$.
At the end of the simulation ($t=350$ms), the probabilities to find an 
atom in state $| 1 \rangle$ and an atom in state $| 2 \rangle$ inside
the trap are $48\%$ and $52\%$, respectively.

In the ``ideal fermionization scenario'', in which the infinitely strongly interacting particles feel the same external potential, the ground state is two-fold degenerate. In our case, this degeneracy is broken since $\mathcal{C}_{| 1 \rangle} \ne \mathcal{C}_{| 2 \rangle}$. Solid and dotted lines in Fig.~\ref{fig_res-rep-ginf}(a) show $|\Psi_{\text{rel}}(z_{12})|^2$,
\begin{eqnarray}
\Psi_{\text{rel}}(z_{12}) = 
\int_{-\infty}^{\infty} 
\Psi(z_1,z_2, t=-t_r) d Z_{\text{CM}},
\label{eq_psirel}
\end{eqnarray}
where $Z_{\text{CM}} = (z_1+z_2)/2$, for the ground state and the first excited state, respectively. 
The difference of the amplitudes for $z_{12}<0$ and $z_{12}>0$
reflects
the asymmetry of the trap potentials (see discussion above).
The ground state wave function is 
greater or equal to zero
everywhere while the first excited state wave function changes sign at $z_{12}= 0$.
The energy difference between the two states is approximately $3\times 10^{-4} E_{\text{ho}}$, corresponding to a time scale of 430ms.
Since parity is not a conserved quantity and since the relative and center-of-mass 
degrees of freedom couple, we expect oscillations between the ground 
state
and 
the 
first excited state at this time scale.
Figure~\ref{fig_res-rep-ginf}(b) shows the normalized overlap 
$\mathcal{O}_{n_{\text{rel}}n_{\text{CM}}}$,
\begin{eqnarray}
\mathcal{O}_{n_{\text{rel}}n_{\text{CM}}}(t) = 
\left|
\frac
{\langle \Psi(z_1,z_2,t) | \phi_{n_{\text{rel}}n_{\text{CM}}}(z_1,z_2)\rangle}
{\sqrt{\langle \Psi(z_1,z_2,t) | \Psi(z_1,z_2,t) \rangle}}
\right|,
\label{eq_overlap}
\end{eqnarray}
between the time-evolving wave packet $\Psi(z_1,z_2,t)$ and the two-body
harmonic oscillator eigenstates $\phi_{n_{\text{rel}}n_{\text{CM}}}(z_1,z_2)$
with trap frequency $\omega_{\text{trap}}$ and relative and 
center-of-mass quantum numbers $n_{\text{rel}}$ and $n_{\text{CM}}$.
The solid line shows the overlap for the anti-symmetric reference wave 
function $\phi_{n_{\text{rel}}n_{\text{CM}}}$ with
$(n_{\text{rel}},n_{\text{CM}})=(1,0)$, which has odd relative parity.
Dotted lines show the
overlaps for states with even relative parity (see figure caption). 
The  
oscillation
period, $T \approx 270$ms, is 
comparable
to but smaller than the 
estimated value of 430ms because the system is modified after $t=-t_r$.
Figure~\ref{fig_res-rep-ginf} demonstrates that two distinguishable particles 
with infinite $g_{1\text{D}}$ but 
$\mathcal{C}_{| 1 \rangle} \ne \mathcal{C}_{| 2 \rangle}$
exhibit unique dynamics that is absent for two identical fermions. 
It could be interesting in future work to tune the system toward and away from the ideal fermionization 
regime
and to explore the resulting dynamics.


\section{Summary and outlook}
\label{sec_summary}

This paper provided a detailed analysis of the two-particle tunneling
dynamics out of an effectively one-dimensional
trap.
Our studies were motivated by experiments by the Heidelberg
group and our analysis was based on full time-dependent simulations
of single- and two-particle systems.
We found that the trap calibration via a WKB analysis leads to an inaccurate
trap parameterization;
this finding is in agreement with a study by 
Lundmark {\it{et al.}}~\cite{Lundmark15}.
Using the reparameterized trapping potential,
our tunneling rates for two identical fermions agree with the 
experimental results for all but 
two magnetic field strengths considered.

Our simulations for the
interacting two-particle systems
made a number of simplifying assumptions.
The dynamics in the tight confinement direction was only incorporated 
indirectly via the renormalized one-dimensional coupling constant. 
For this, a harmonic trap in the tight direction was assumed.
Moreover, we assumed simple short-range or zero-range interaction
potentials.
Deep-lying bound states and coupled channel effects were neglected entirely.
Using the renormalized one-dimensional
coupling constant $g_{1{\text{D}}}$ with the transverse
frequency $\sqrt{p(0)}\omega_{\rho}^{\text{ref}}$ 
as input,
our simulations reproduced the upper branch tunneling dynamics of the 
interacting two-particle system reasonably well.
Our simulation results 
for
the molecular branch dynamics agreed with the overall trend
of the experiment but did not yield quantitative agreement.
We argued that the actual transverse confinement felt by the atoms
in the presence of the magnetic field gradient may
be weaker than in the absence of the magnetic field gradient.
This motivated us to calculate the one-dimensional coupling constant
using a weaker transverse trapping frequency as input.
The resulting two-particle tunneling rates are in agreement
with the experimentally measured rates over the entire range of 
magnetic field strengths considered.
We note that our finding is consistent with Ref.~\cite{wall},
which found that the non-separability of a gaussian trap
affects the tunneling rate in a double-well geometry.

Our work suggests a number of
follow-up studies.
It would be interesting to extend the dynamical simulations
to more particles
and/or to include the tight confining directions.
It would also be interesting to
prepare other initial one- and two-particle states.
For example, it would be interesting to investigate the
tunneling dynamics from initial excited metastable states.

\section{Acknowledgement}
We thank S.~Jochim and G.~Z\"urn for clarifying various aspects of the
experimental protocols  and for stimulating discussions.
We also thank P.~Zhang for discussions,
and Yangqian~Yan and X.~Y.~Yin for generous help with parallelizing 
our time evolution code.
Some of the calculations were performed on the WSU HPC.
Support by the NSF through grant 1415112 is gratefully acknowledged.

\appendix

\section{State dependence of the trapping potential and Breit-Rabi formula}
\label{app_BR}
We consider an atom with total (orbital and spin) electronic angular momentum 
quantum number $J=1/2$ and 
nuclear spin $I$ ($I=1$ for $^6$Li). 
In the absense of an external magnetic field, the energy difference $\Delta W$
between the hyperfine states $|F=I-1/2,m_F \rangle$ and $|F=I+1/2,m_F \rangle$ 
is independent of $m_F$.
For $^6$Li with $|F=1/2 \rangle$ and $|F=3/2 \rangle$, 
$\Delta W$ is equal to $228.205$ MHz~\cite{laser-cooling}.
According to the Breit-Rabi formula~\cite{Breit-Rabi,Ramsey-56},
the energy $W^{\text{BR}}_{|F,m_F \rangle}(B)$ of the hyperfine 
state $|F,m_F \rangle$ 
in an external magnetic field of strength $B$ is
\begin{eqnarray}
W^{\text{BR}}_{|F,m_F \rangle}(B) = 
- \frac{\Delta W}{2(2 I + 1)} 
+ g_I \mu_B m_F B \nonumber \\
\pm \frac{\Delta W}{2}
\left(1+ \frac{4 m_F}{2 I + 1}x + x^2 \right)^{1/2},
\label{Eq_BR}
\end{eqnarray} 
where $x = (g_J - g_I) \mu_B B /\Delta W $, 
$g_J$ is the Land\'e factor, 
and $g_I$ characterizes the magnetic moment of the nucleus.
The plus and minus signs refer to states $F=I+1/2$ and $F=I-1/2$, respectively.
The constants $g_J = 2.0023019(24)$ and $g_I = -0.0004476493(45)$ are 
determined experimentally~\cite{Boklen}.
Figure~\ref{fig_BR-energy}
shows the magnetic field dependence of the hyperfine states of 
$^6$Li for $F=1/2$ and $F=3/2$. 
The slope of these energy curves equals the negative of the magnetic moment of the atom~\cite{Ramsey-56},
yielding
\begin{eqnarray}
c_{| F,m_F \rangle}^{\text{BR}}(B) = -
\frac{1}{\mu_B} 
\frac{d~ }{d B} W^{\text{BR}}_{|F,m_F \rangle}(B).
\label{eq_c}
\end{eqnarray} 
Equation~(\ref{eq_c}) characterizes the state and magnetic field 
dependence of the trapping potential
(see Sec.~\ref{sec_system} of the main text).
The coefficients calculated according to Eqs.~(\ref{Eq_BR}) and (\ref{eq_c})
are referred to as Breit-Rabi coefficients in the main text. 

\section{Time dynamics for two identical fermions in an anti-symmetric state and in a product state}
\label{app_sup-vs-pro}

The wave packet of two identical fermions
is anti-symmetric under the exchange of the particles. 
To calibrate the trap
(see Sec.~\ref{sec_upper-trp-cal}), the assumption in using the WKB approximation was that the dynamics could be described as if a single particle was tunneling out of the first excited trap state. Our numerical simulations show that the tunneling rates are, indeed, very similar. For the parameters listed in the third row of 
Table~\ref{table-used-param-upper-NI}, we find $\gamma = 13.8 \text{s}^{-1}$ for the two-particle system and $\gamma_{\text{sp}} = 13.5 \text{s}^{-1}$ for the single-particle system. As we discuss now, the tunneling dynamics is, however, quite different.

\begin{figure}[htbp]
\centering
\includegraphics[angle=0,width=0.4\textwidth, clip=true]{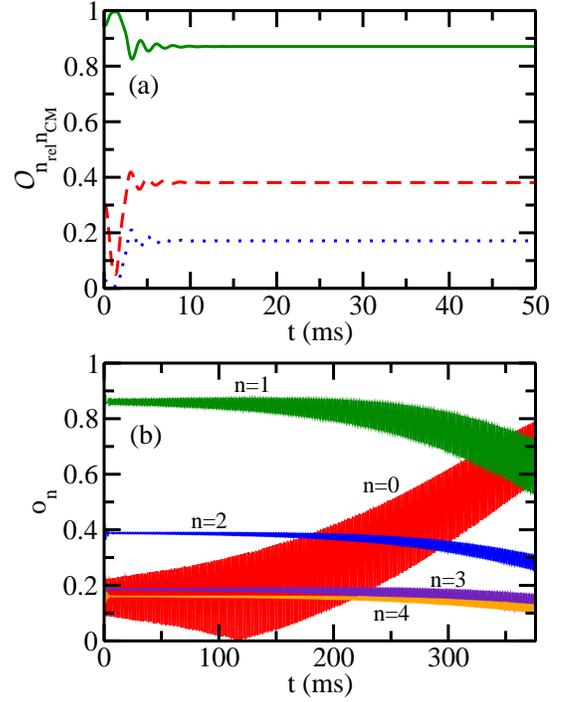}
\caption{(Color online)
(a)~Analysis of the time dynamics for two identical fermions.
The solid, dashed, and dotted lines show the normalized overlap $\mathcal{O}_{n_{\text{rel}}n_{\text{CM}}}$ between the wave packet $\Psi(z_1,z_2,t)$ and the harmonic oscillator states with odd relative parity [from top to bottom, $(n_{\text{rel}},n_{\text{CM}}) =$ (1,0), (1,1), and (1,2)].
(b)~Analysis of the time dynamics for a single atom in state $|2\rangle$, prepared in the first excited trap state at $t = -t_r$.
The lines show the normalized overlaps $o_n(t)$ between the wave packet $\Psi(z,t)$ and the harmonic 
oscillator states with $n=0-4$.
For both panels, the parameters $p(t=0) = 0.68$,
$z_{\text{R}} = 8.548 a_{\text{ho}}$,
$c_{| 2 \rangle}^{\text{BR}} = 0.99601$,
and $B'=1890$G/m are used.
}
\label{fig_sup-vs-pro}
\end{figure}

Figure~\ref{fig_sup-vs-pro}(a)
shows the normalized overlaps $\mathcal{O}_{n_{\text{rel}}n_{\text{CM}}}$ 
[see Eq.~(\ref{eq_overlap})]
between the time-evolving anti-symmetric two-particle wave packet $\Psi(z_1,z_2,t)$ 
and the two-body
harmonic oscillator eigenstates $\phi_{n_{\text{rel}}n_{\text{CM}}}(z_1,z_2)$
with trap frequency $\omega_{\text{trap}}$ and relative and center-of-mass 
quantum numbers $n_{\text{rel}}$ and $n_{\text{CM}}$.
The solid, dashed and dotted lines show the overlaps for $n_{\text{rel}} = 1$
and $n_{\text{CM}} = 0, 1$, and $2$, respectively.
The normalized overlaps oscillate for a short time
($t \lesssim 10$ms) 
and quickly approach constants. We see essentially constant overlaps till the 
end of our simulation at $t = 500$ms.
This indicates that the shape of the wave packet in region $R_2$ is
constant in time.
The overlaps vanish for even $n_{\text{rel}}$, indicating that the 
anti-symmetry of the wave packet is preserved during the time evolution.

Figure~\ref{fig_sup-vs-pro}(b)~shows
the normalized overlap $o_n(t)$,
\begin{eqnarray}
o_n(t) = 
\left|
\frac{\langle \Psi(z,t) | \phi_n(z)\rangle}
{\sqrt{\langle \Psi(z,t) | \Psi(z,t) \rangle}} \right|,
\label{eq_ov}
\end{eqnarray} 
between the time-dependent single-particle wave packet
$\Psi(z,t)$ and the
time-independent single-particle harmonic oscillator functions $\phi_n(z)$
with quantum number $n$,
$n=0-4$.
The overlaps shown in Fig.~\ref{fig_sup-vs-pro}(b)~oscillate at a frequency
that is close to the natural trap frequency $\omega_{\text{trap}}$ for $t>20$ms.
Moreover, the ``envelopes'' of the overlaps change in time, indicating  
that the shape of the wave packet in region $R_2$ changes with
time.
At $t=0$, the wave packet has a finite overlap with the ground state 
harmonic oscillator wave function due to the change of the trapping potential.
The contribution of the harmonic oscillator ground state to the wave packet 
is almost constant in time while the contributions of higher energy states 
deplete.
This results in the increase of the normalized overlap $o_0(t)$
[see Fig.~\ref{fig_sup-vs-pro}(b)].
In other words, as $P_{\text{sp,in}} (t)$ decreases, the wave packet looks more like the lowest-lying trap state as opposed to the initial state.
As a consequence, the decay of $P_{\text{sp,in}} (t)$ with time
deviates slightly from an exponential.

\section{Time propagation via Chebyshev expansion}
\label{app_Cheb}

The time evolution of the two-particle wave packet $\Psi(z_1,z_2,t)$ 
is given by 
\begin{eqnarray}
\Psi(z_1,z_2,t) = \mathcal{U} (t-t_0) \Psi(z_1,z_2,t_0),
\label{eq_time-evol}
\end{eqnarray} 
where the time-evolution operator $\mathcal{U} (t-t_0)$ is
\begin{eqnarray}
\mathcal{U} (t-t_0) = \exp[-i H (t-t_0) / \hbar] .
\label{eq_time-evol-op}
\end{eqnarray} 
To evaluate Eq.~(\ref{eq_time-evol}), one has to expand the time-evolution 
operator $\mathcal{U} (t-t_0)$ 
in powers of $-i H (t-t_0) / \hbar$.
It has been shown that expanding $\mathcal{U} (t-t_0)$ in terms of 
the complex Chebyshev polynomials $\phi_k$,
\begin{eqnarray}
\mathcal{U} (t-t_0) = \sum_{k=0}^N a_k 
\phi_k \left(\frac{-i H (t-t_0)}{\hbar R}\right),
\label{eq_Cheb}
\end{eqnarray} 
provides an efficient means to determine
the time evolution of the wave packet~\cite{Tal-Ezer_re}.
Here, $R$ is a real and positive number that has been introduced to normalize the 
argument of $\phi_k$ such that $ -i H (t-t_0)/(\hbar R)\in [-i,i]$.
A key point is that the recursion relation
\begin{eqnarray}
\phi_k(X) = 2 X \phi_{k-1}(X) + \phi_{k-2}(X)
\label{eq_rec-Cheb}
\end{eqnarray} 
for the $k^{\text{th}}$ Chebyshev polynomial
enables one to readily reach high orders in the expansion, allowing one to 
go to large $N$ in Eq.~(\ref{eq_Cheb}) and, correspondingly, to large $t-t_0$.
The expansion coefficients $a_k$ are expressed in terms of Bessel functions
of the first kind of order $k$. 
For more details, the reader is referred to 
Refs.~\cite{Tal-Ezer_re, Leforestier91}.

We use this approach to evaluate $\Psi(z_1,z_2,t)$ for $t>-t_r$.
We choose time steps
around $0.2 \omega^{-1}$ and $N$ up to $800$.
A time step of $0.2 \omega^{-1}$ is small enough to resolve the tunneling 
dynamics and to extract the time dependence of the flux reliably, i.e., at the
few percent accuracy level.

\section{Preparation of the initial state}
\label{app_im-time}

In Secs.~\ref{sec_molecular} and~\ref{sec_upper} we need the initial 
(equilibrium) state of the trapped particles at $t=-t_r$.
We use $p(t=-t_r)=0.795$ for all cases. 
For this trap depth, tunneling 
is highly suppressed and the system is in a metastable state, i.e.,
it has a lifetime much larger 
than the time scale of the forthcoming tunneling process.
To prepare the initial state, we ``artificially'' 
put a hard wall at $z = 11 a_{\text{ho}}$ 
(the top of the barrier is located at $z \approx 9 a_{\text{ho}}$).
We changed the position of the hard wall somewhat without seeing a 
notable change in the results.
For example, for the upper branch calculations at $B=900$G, we changed the
position of the hard wall to $10a_{\text{ho}}$ and $12a_{\text{ho}}$
and found that the overlap between the resulting initial states and the
state prepared with the hardwall at $11a_{\text{ho}}$ deviated from
1 by less than $10^{-6}$.
This artificial boundary condition leaves the trap in the ``inside'' region 
unchanged and completely turns off the tunneling. 
The resulting eigenstates are to a very good approximation equal to
the metastable states of the trap with finite barrier.
 
We start with an initial wave packet that has a finite overlap with
the state that we are looking for
and act with the time-evolution operator, Eq.~(\ref{eq_time-evol-op})
with imaginary time $\tau$, on the initial wave 
packet~\cite{Davies_im, Tal-Ezer_im}.
To propagate the wave packet in imaginary time, we use the real time-propagation 
methods discussed in Appendices~\ref{app_Cheb} and~\ref{app_ZR-time} with 
$t$ replaced by $\tau / i$.
The initial wave packet can, in principle, be expanded in terms of 
unknown eigenfunctions of the Hamiltonian.
After application of the time-evolution operator with imaginary time, each 
term in the expansion
gets damped at a rate that is proportional to its energy.
Thus, states with high energy decay fastest and eventually only the lowest 
energy state survives.
We perform the imaginary time propagation using small $\Delta \tau$ and
normalize the wave packet to one after each step.
This process can be generalized to find excited states by removing the 
lower energy eigenstates from the Hilbert space~\cite{Tal-Ezer_im}.
In practice, this is done by orthogonalizing the evolving 
wave packet and the lower energy eigenstate(s) after each time step.

To implement the Chebychev expansion based approach with imaginary time, 
we expand the exponential function in terms of real Chebychev 
polynomials and use the corresponding recursion relation~\cite{Tal-Ezer_im}.
We typically use about 15 terms in the series and time steps around 
$(0.005 \omega^{-1})/i$.
The Trotter formula based propagation scheme with imaginary time does not 
involve integrals over highly oscillatory functions and the calculations 
are computationally
much less expensive than those for the real time propagation.


\section{Time propagation for Hamiltonian with two-body zero-range interaction}
\label{app_ZR-time}

The time propagation based on the Chebychev expansion is not applicable to the 
two-particle Hamiltonian with two-body zero-range interaction.
In this case, we use a propagator that accounts for the two-body zero-range 
interaction exactly to determine the time evolution 
of the wave packet~\cite{Blinder_88, Wodkiewicz_91}. 
This propagator has recently been used in Monte Carlo simulations for 
systems with zero-range interactions~\cite{Yan_15}.
This appendix summarizes our implementation of the  real
time evolution in the presence of a zero-range two-body potential.
The wave packet $\Psi$ at time $t+\Delta t$ can be written as
\begin{eqnarray}
\Psi(z_1,z_2,t + \Delta t) = \nonumber \\
\int_{-\infty}^{\infty}\int_{-\infty}^{\infty} 
\rho (z'_1,z'_2; z_1,z_2; \Delta t) \Psi(z'_1,z'_2,t)dz'_1 dz'_2,
\label{eq_time-evol-ZR}
\end{eqnarray} 
where the zero-range propagator $\rho$ is defined through
\begin{eqnarray}
\rho (z'_1,z'_2; z_1,z_2; \Delta t) = \nonumber \\
\langle z'_1,z'_2|\exp(-i H \Delta t / \hbar)|z_1,z_2\rangle.
\label{eq_propagator1}
\end{eqnarray} 
In free space, i.e., when the two-body Hamiltonian consists of the kinetic
energy and the zero-range interaction, the propagator 
$\rho_{\text{free}}$ can be written as
\begin{eqnarray}
\rho_{\text{free}} (z'_1,z'_2; z_1,z_2; \Delta t) = 
\rho_{\text{free}}^{\text{sp}} (z'_1, z_1, \Delta t) 
\nonumber \\ \times 
\rho_{\text{free}}^{\text{sp}} (z'_2, z_2, \Delta t) 
\rho_{\text{free}}^{\text{rel}} (z_1' - z'_2, z_1 - z_2, \Delta t),
\label{eq_propagator2}
\end{eqnarray} 
where $\rho_{\text{free}}^{\text{sp}}$,
\begin{eqnarray}
\rho_{\text{free}}^{\text{sp}} (z', z, \Delta t) = 
\nonumber \\ 
\left( \frac{m}{2 \pi i \Delta t \hbar} \right)^{1/2}
\exp\left( -\frac{m (z-z')^2}{2 i \Delta t \hbar} \right),
\label{eq_propagator3}
\end{eqnarray} 
accounts for the single-particle kinetic energy
and $\rho_{\text{free}}^{\text{rel}}$,
\begin{eqnarray}
\rho_{\text{free}}^{\text{rel}} (z', z, \Delta t) =
1-
\exp\left( -\frac{m (z z' + |z z'|)}{2 i \Delta t \hbar} \right) 
\nonumber \\ \times 
\sqrt{\frac{m i \pi \Delta t}{4 \hbar}}
\frac{g_{1\text{D}}}{\hbar}
\exp\left( u^2 \right) \text{erfc}(u),
\label{eq_propagator4}
\end{eqnarray} 
for the two-body zero-range potential~\cite{Blinder_88, Wodkiewicz_91}.
In Eq.~(\ref{eq_propagator4}), erfc denotes the complementary error function
and $u$ is equal to 
$m(|z|+|z'|+ i g_{1\text{D}} \Delta t / \hbar) / \sqrt{4 m i \Delta t \hbar}$.
For infinitely strong interaction, i.e., for $|g_{1\text{D}}| = \infty$, 
Eq.~(\ref{eq_propagator4})
simplifies to 
\[  \rho_{\text{free}}^{\text{rel}} (z', z, \Delta t) = \left\{
\begin{array}{l r}
  1 - \exp\left( -\frac{m z z'}{ i \Delta t \hbar} \right) &
  \text{for} \quad z z' > 0
  \\
  0 &
  \text{for} \quad z z' \le 0
\end{array} \right. . \] 
In the presence of the external potential $V_{\text{ext}}$, 
we use the Trotter formula~\cite{Trotter59},
\begin{eqnarray}
\rho (z'_1,z'_2; z_1,z_2; \Delta t) \approx 
\exp\left(-\frac{i \Delta t}{2 \hbar} V_{\text{ext}}(z_1', z_2') \right)  
\nonumber \\ \times 
\rho_{\text{free}} (z'_1,z'_2; z_1,z_2; \Delta t)
\exp\left(-\frac{i \Delta t}{2 \hbar} V_{\text{ext}}(z_1,  z_2 ) \right).
\label{eq_propagator5}
\end{eqnarray} 
This decomposition yields an error in the propagator that is proportional 
to $\Delta t^3$.
We use Eq.~(\ref{eq_time-evol-ZR}) with $\rho$ given by 
Eq.~(\ref{eq_propagator5}) to propagate the wave packet in real time 
for each time step $\Delta t$. 
Unlike the Chebychev expansion approach, the Trotter formula based approach
is limited to small $\Delta t$. 
Importantly, the integrand in Eq.~(\ref{eq_time-evol-ZR}) 
oscillates with a frequency that is proportional to $1/\Delta t$.
To resolve these oscillations we need to choose a sufficiently dense
spatial grid for the numerical integration of the right hand side of
Eq.~(\ref{eq_time-evol-ZR}). 
We typically use a grid spacing 
$\Delta z_j / a_{\text{ho}} \le \Delta t / (10 \omega^{-1})$ ($j=1$ and 2).
We find that a value of $\Delta t \leq 0.2 \omega^{-1}$ ensures that the
norm of the wave packet, accounting for the absorbed portion of the wave 
packet, is 0.99 (or even closer to one) at the end of our simulation.
Due to the need to evaluate the two-dimensional integral for each grid point, 
the Trotter formula based propagation scheme is much more computationally
demanding than the Chebychev polynomial based propagation scheme.

\section{Application of the absorbing potential}
\label{app_damping}

The damping of the wave packet in the numerical regions $R_{1An}$, $R_{0n}$,
and $R_{1Bn}$ has the same effect as an absorbing potential.
After each time step, we multiply the wave packet
by $\mathcal{D}(z_1) \mathcal{D}(z_2)$~\cite{Sun09}, where
\[ \mathcal{D}(z) = \left\{
\begin{array}{l l}
  1 &
  \text{for} \quad z<z_{d}
  \\
  \exp{\left[-\alpha \left(\frac{z-z_{d}}{\Delta_d}\right)^{n_d}\right]}&
  \text{for} \quad  z \ge z_{d}
\end{array} \right. . \] 
Here, $\Delta_d$, $\alpha$, and $n_d$ are parameters whose values 
depend, in general, on the kinetic energy of the particle that is being 
absorbed.
We use $\Delta_d = 10 a_{\text{ho}}$, $\alpha = 5$, and $n_d=2$ with 
$z_{hw}-z_{d} \ge 6 a_{\text{ho}}$, where $z_{hw}$ is the position of the hard wall at the end of the simulation grid.
This parameter combination ensures that the reflection from the end of the
numerical box is negligibly small.

\section{Flux analysis}
\label{app_flux-anal}

In this Appendix we discuss how to extract physical quantities from 
the density flux.
$P_n(t)$ denotes the probability to find $n$ particles ($n=0,1,$ or 2) inside 
the trap at time $t$.
$P_2(t)$ is obtained by integrating the density 
$|\Psi(z_1,z_2,t)|^2 $
over the region $R_2$ (see Fig.~\ref{fig_regions}),
\begin{eqnarray}
P_2(t) = \int_{R_2} |\Psi(z_1,z_2,t)|^2 dz_1 dz_2.
\end{eqnarray}
The initial condition is given by $P_2(-t_r) = 1$, i.e., at time $t=-t_r$ 
both particles are inside the trap.
For $t > -t_r$, we have
$P_2(t) + P_1(t) + P_0(t) = 1$.
The density $|\Psi(z_1,z_2,t)|^2 $ can flow from one region to another
during the time propagation.
To quantify the change of $P_n(t)$, we use the current ${\bf{j}}(z_1, z_2,t)$,
\begin{eqnarray}
{\bf{j}}(z_1, z_2,t) = 
-\frac{\hbar}{m} 
\text{Im}
\left[ \Psi^*(z_1,z_2,t) \; {\boldsymbol{\nabla}} \Psi(z_1,z_2,t)
\right],
\end{eqnarray}
where 
\begin{eqnarray}
{\boldsymbol{\nabla}} = 
\frac{\partial}{\partial z_1} {\bf{\hat{z}}}_1 +
\frac{\partial}{\partial z_2} {\bf{\hat{z}}}_2.
\end{eqnarray}
Here, ${\bf{\hat{z}}}_1$ and ${\bf{\hat{z}}}_2$ are the unit vectors 
in the $z_1$ and $z_2$ directions, respectively.
At each point in time and space, the continuity equation requires
\begin{eqnarray}
\frac{\partial |\Psi(z_1,z_2,t)|^2}{\partial t}
+ {\boldsymbol{\nabla}} \cdot {\bf{j}}(z_1, z_2,t) = 0.
\label{eq_cont}
\end{eqnarray}
If we integrate Eq.~(\ref{eq_cont}) over the region $R_i$ 
($R_i$ can be equal to $R_2$, $R_{1A}$, or $R_{1B}$; see Fig.~\ref{fig_regions}),
we find
\begin{eqnarray}
\frac{\partial}{\partial t} \int_{R_i} |\Psi(z_1,z_2,t)|^2 \; dz_1 dz_2 =
\nonumber \\
- \int_{R_i} {\boldsymbol{\nabla}} \cdot {\bf{j}}(z_1, z_2,t) \; dz_1 dz_2.
\label{eq_int-j-1}
\end{eqnarray}
The left hand side of Eq.~(\ref{eq_int-j-1}) is the rate at which the 
probability of finding the system in region $R_i$ changes.
To simplify the right hand side, we use the divergence theorem
in two spatial dimensions,
\begin{eqnarray}
\int_{R_i} {\boldsymbol{\nabla}} \cdot {\bf{j}}(z_1, z_2,t) \; dz_1 dz_2 =
\nonumber \\ \oint_{B_i} {\bf{j}}(z_1, z_2,t) \cdot {\bf{\hat{n}}}_i \; d l.
\end{eqnarray}
Here, $d l$ is the line element corresponding to the closed 
boundary $B_i$ that encircles region $R_i$ and
${\bf{\hat{n}}}_i$ is the unit vector perpendicular to the 
boundary and directed out of the region $R_i$.
Equation~(\ref{eq_int-j-1}) can thus be written as
\begin{eqnarray}
\frac{\partial}{\partial t} \int_{R_i} |\Psi(z_1,z_2,t)|^2  \; dz_1 dz_2 =
\nonumber \\
- \oint_{B_i} {\bf{j}}(z_1, z_2,t) \cdot {\bf{\hat{n}}}_i \; d l.
\label{eq_int-j-2}
\end{eqnarray}
The change of the probability to find the system in region $R_i$
can be obtained from the flux through the boundary $B_i$.
Applying Eq.~(\ref{eq_int-j-2}) to region $R_2$, we obtain
\begin{eqnarray}
\frac{\partial P_2(t)}{\partial t}  =
- \oint_{B_2} {\bf{j}}(z_1, z_2,t) \cdot {\bf{\hat{n}}}_2 \; d l
\end{eqnarray}
or
\begin{eqnarray}
P_2(t) = 
1
- \int_{-t_r}^{t}
\oint_{B_2} {\bf{j}}(z_1, z_2,t) \cdot {\bf{\hat{n}}}_2  \; d l \; d t.
\label{eq_int-j-R2}
\end{eqnarray}

To extract additional information from Eq.~(\ref{eq_int-j-2}), we break 
the boundary $B_2$ into pieces.
In particular, flux through the boundary $b_{2,0}$ corresponds to the 
correlated tunneling of two particles (pair tunneling) and flux through 
the boundaries $b_{2,1A}$ and $b_{2,1B}$ corresponds to single-particle 
tunneling (one particle tunnels and one remains in the trap). 
To quantify this in terms of tunneling rates, we define
the rate $\gamma_2$ at which $P_2(t)$ decays during
the time $\Delta t$ through
\begin{eqnarray}
\gamma_2 =
- \frac{1}{P_2(t)} \frac{\Delta P_2(t)}{\Delta t}.
\end{eqnarray} 
Next, we divide the quantity $\Delta P_2(t)$ into two pieces, namely the 
change $\Delta P_{2 \to 0}(t)$ due to the pair tunneling (flux through 
the boundary $b_{2,0}$) 
and the change $\Delta P_{2 \to 1}(t)$ due to the single-particle tunneling  
(flux through the boundaries $b_{2,1A}$ and $b_{2,1B}$),
\begin{eqnarray}
\Delta P_2(t) = \Delta P_{2 \to 0}(t) + \Delta P_{2 \to 1}(t).
\end{eqnarray} 
Defining the pair tunneling rate $\gamma_{\text{P}}$ and the single-particle 
tunneling rate $\gamma_{\text{s}}$,
\begin{eqnarray}
\gamma_{\text{P}} =
- \frac{1}{P_2(t)} \frac{\Delta P_{2 \to 0}(t)}{\Delta t}
\end{eqnarray} 
and
\begin{eqnarray}
\gamma_{\text{s}} =
- \frac{1}{P_2(t)} \frac{\Delta P_{2 \to 1}(t)}{\Delta t},
\end{eqnarray} 
we have $\gamma_2 = \gamma_{\text{P}} + \gamma_{\text{s}}$.
$\gamma_{\text{P}}$ and $\gamma_{\text{s}}$ oscillate in time for $t$ not much larger than
$t_r$ (typically $t \lesssim 20$ms) and are essentially constant for large 
$t$ ($t \gtrsim 20$ms).
The values reported in the main text are obtained by fitting the numerical 
data for sufficiently large $t$.

If 
$z_{\text{d}} \to \infty$, we can find $P_1(t)$ by
integrating the density $|\Psi(z_1,z_2,t)|^2 $ over the regions $R_{1A}$ 
and $R_{1B}$ (see Fig.~\ref{fig_regions}) or, equivalently, by analyzing
the flux through boundaries $b_{2,1A}$, $b_{1A,0}$, $b_{2,1B}$, and $b_{1B,0}$.
The average direction of the flux is into the region $R_{1A}$ ($R_{1B}$) 
through boundary $b_{2,1A}$ ($b_{2,1B}$) and out of the region $R_{1A}$ 
($R_{1B}$) through boundary $b_{1A,0}$ ($b_{1B,0}$).
In the upper branch simulations, we find vanishing flux through boundaries
$b_{1A,0}$ and $b_{1B,0}$.
Thus, without worrying about the finite size of the simulation box, we can determine
$P_1(t)$ as the sum of the fluxes through boundaries $b_{2,1A}$ and $b_{2,1B}$,
\begin{eqnarray}
P_1(t) = 
- \int_{-t_r}^{t}
\bigg\{
\int_{b_{2,1A}} {\bf{j}}(z_1, z_2,t) \cdot {\bf{\hat{n}}}_{1A}  \; d l
~~~~~ \nonumber \\ 
+
\int_{b_{2,1B}} {\bf{j}}(z_1, z_2,t) \cdot {\bf{\hat{n}}}_{1B}  \; d l 
\bigg\}
d t.
\label{eq_int-j-R1}
\end{eqnarray}
It should be noted that if the flux through boundaries $b_{1A,0}$ and 
$b_{1B,0}$ is non-zero, then the evaluation of $P_1(t)$ is more involved; 
this case is not discussed here.

\section{Additional comments on the WKB approximation}
\label{appendix_wkb}
As discussed in the main text,
the WKB approximation yields
single-particle
tunneling rates that are smaller 
(larger) than the exact tunneling rates
for the 
trap ground state 
(first excited trap state). 
To elaborate on this behavior, 
we determine $p(t=0)$ for the trap ground state, the first
excited trap state, and the second excited trap state
such that (a) ${\mathcal{T}}=0.06267$ and 
(b) ${\mathcal{T}}=0.0063$.
We then perform exact single-particle
time propagation calculations
for these cases, starting with a quasi-eigenstate
(either the ground state, the first excited trap state, or the
second excited trap state)
for $p(t=-t_r)=0.795$.
Table~\ref{table_app_WKB} summarizes the resulting tunneling
rates $\gamma_{\text{sp}}^{\text{num}}$. It can be seen that
$\gamma_{\text{sp}}^{\text{num}}$ is approximately independent
of the state number but, as expected, strongly dependent on the actual
barrier the particle has to tunnel through.
\begin{table*}
\caption{Single-particle WKB versus exact
tunneling rates. 
The tunneling coefficients for cases (a)
and (b) are
$\mathcal{T} = 0.06267$ and 
$\mathcal{T} = 0.0063$, respectively.
The third column reports the value of $p(t=0)$ 
for which the trap ground state, first excited trap state, and 
second excited trap state yield the desired $\mathcal{T}$.
The fourth and fifth columns report the WKB frequency
$f^{\text{WKB}}$ and the single particle WKB tunneling rate
$\gamma_{\text{sp}}^{\text{WKB}}$, Eq.~(\ref{eq_WKB-rate}),
respectively.
For comparison, the sixth column reports 
the tunneling rate $\gamma_{\text{sp}}^{\text{num}}$
obtained from our exact time-dependent simulations.
The calculations are performed for $\mathcal{C}=1890$G/m, 
$V_0 = 56.16 E_{\text{ho}}$, and $z_{\text{R}} = 8.548 a_{\text{ho}}$.
}
\begin{tabular}{cccccc}
case &
trap state &
$p(t=0)$ & 
$f^{\text{WKB}}$ ($\text{ms}^{-1}$) &
$\gamma^{\text{WKB}}_{\text{sp}}$ ($\text{ms}^{-1}$) &
$\gamma^{\text{num}}_{\text{sp}}$ ($\text{ms}^{-1}$)
\\
\hline
\hline
(a) &
gr. st &
0.63540 & 
0.322 &
0.0202 &
0.0330
\\
(a) &
1st exc. st. &
0.67687& 
1.104 &
0.0692 &
0.0330
\\
(a) &
2nd exc. st. &
0.71486 &
1.999 &
0.1253 &
0.0329
\\
\hline
(b) &
gr. st &
0.6489 & 
0.352 &
0.00222 &
0.00406
\\
(b) &
1st exc. st. &
0.6899& 
1.1732 &
0.00739 &
0.00403
\\
(b) &
2nd exc. st. &
0.7277 &
2.0965 &
0.013215 &
0.00400
\end{tabular}
\label{table_app_WKB}
\end{table*}
Due to the dependence of
$f^{\text{WKB}}$ on the state
(through the WKB energy),
the WKB rates 
$\gamma_{\text{sp}}^{\text{WKB}}$ for the three
states 
vary by about a factor of 6
for cases (a) and (b).
For the parameters considered
in Table~\ref{table_app_WKB} and in the
main text, the WKB rate for the ground state 
is smaller than
that obtained through the full time propagation, with the ratio
$\gamma_{\text{sp}}^{\text{WKB}}/\gamma_{\text{sp}}^{\text{num}}$
depending on the exact shape of the trap.
For the excited states, in contrast,
the WKB rates are larger than
those obtained through the full time propagation.



\begin{thebibliography}{49}
\bibitem [{footnote6()}]{footnote6}%
  \BibitemOpen
  \bibinfo {note} {The size of the free-space molecule for the zero-range
  interaction is equal to $a_{1\text{D}}$ but for the finite-range interaction
  corrections must be applied.}\BibitemShut {Stop}%
\bibitem [{footnote5()}]{footnote5}%
  \BibitemOpen
  \bibinfo {note} {The experimental data for the interacting systems are taken
  from Table II of Ref.~\cite{Jochim4} and the experimental data point for the
  non-interacting system is obtained by adding the measured single-particle
  tunneling rates reported in Table I of Ref.~\cite{Jochim4}.}\BibitemShut
  {Stop}%
\end{thebibliography}

%

\end{document}